\documentclass[]{aastex}
\newcommand{\lta}{$\; \buildrel < \over \sim \;$}
\newcommand{\simlt}{\lower.5ex\hbox{\lta}}
\newcommand{\gta}{$\; \buildrel > \over \sim \;$}
\newcommand{\simgt}{\lower.5ex\hbox{\gta}}

\newcommand{\kms}{{\rm\,km\,s^{-1}}}
\newcommand{\kpc}{{\rm\,kpc}}
\newcommand{\mpc}{{\rm\,Mpc}}
\newcommand{\Mpc}{{\rm\,Mpc}}
\newcommand{\msun}{{\rm\,M_\odot}}

\newcommand{\cm}{{\rm\,cm}}
\newcommand{\Gyr}{{\rm\,Gyr}}

\newcommand{\clus}{{\it Cool+SF}}
\newcommand{\adia}{{\it Adiabatic}}
\newcommand{\lowr}{{\it Lowres}}
\newcommand{\dark}{{\it Dark}}
\newcommand{\hire}{{\it Hires}}

\newcommand{\SZ}{Sunyaev-Zel'dovich}

\slugcomment{Submitted to the Astrophysical Journal}
\shortauthors{G. F. Lewis et. al.}
\shorttitle{Physics and Resolution in Cluster Simulations}
\begin{document}

\title{  The Effects  of Gas  Dynamics, Cooling,  Star  Formation, and
Numerical   Resolution   in  Simulations   of   Cluster  Formation   }

\author{
Geraint F.  Lewis\altaffilmark{1}, 
Arif Babul\altaffilmark{2},
Neal   Katz\altaffilmark{3},    
Thomas   Quinn\altaffilmark{4},   
Lars Hernquist\altaffilmark{5} \& 
David H. Weinberg\altaffilmark{6}}

\altaffiltext{1}{ Fellow of the Pacific Institute of Mathematical
Sciences 1998-1999, \\ Dept. of Physics and Astronomy, University of
Victoria, PO Box 3055, Victoria, B.C., V8W 3P6, Canada \\ \&
Astronomy Dept., University of Washington, Box 351580, Seattle WA
98195-1580, U.S.A.  \\ Electronic mail: {\tt
gfl@astro.washington.edu}\\ Electronic mail: {\tt
gfl@uvastro.phys.uvic.ca} }

\altaffiltext{2}{
Dept. of Physics and Astronomy, University of Victoria, 
PO Box 3055,  Victoria, B.C.,  V8W 3P6, Canada
\\
Electronic mail: {\tt babul@uvic.ca}}

\altaffiltext{3}{
Dept. of Physics and Astronomy, University of Massachusetts, 
Amherst,  MA 01003, U.S.A.
\\
Electronic mail: {\tt nsk@kaka.phast.umass.edu}}

\altaffiltext{4}{
Astronomy Dept., University of Washington, Box 351580, 
Seattle WA 98195-1580, U.S.A.
\\
Electronic mail: {\tt trq@astro.washington.edu}}

\altaffiltext{5}{
Department of Astronomy, Harvard University, 60 Garden
Street, Cambridge MA 02138
\\
Electronic mail: {\tt lars@cfa.harvard.edu}}

\altaffiltext{6}{
Dept. of Astronomy, The Ohio State University, Columbus, 
OH 43210, U.S.A.
\\
Electronic mail: {\tt dhw@astronomy.ohio-state.edu}}

\begin{abstract}
We present  the analysis  of a  suite of simulations  of a  Virgo mass
galaxy cluster. Undertaken within  the framework of standard cold dark
matter  cosmology,  these  simulations  were  performed  at  differing
resolutions and with increasingly complex physical processes, with the
goal  of identifying  the  effects of  each  on the  evolution of  the
cluster.  We  focus on  the cluster at  the present epoch  and examine
properties including the radial distributions of density, temperature,
entropy and  velocity.  We also map  `observable' projected properties
such as  the surface mass  density, X-ray surface brightness  and \SZ\
signature.    We   identify   significant  differences   between   the
simulations,  which highlights  the  need for  caution when  comparing
numerical  simulations  to  observations  of galaxy  clusters.   While
resolution  affects   the  inner   density  profile  in   dark  matter
simulations, the addition of  a gaseous component, especially one that
cools and  forms stars, affects  the entire cluster.  For  example, in
simulations  with   gas  dynamics   but  no  cooling,   improving  the
gravitational force  resolution from 200  kpc to 14 kpc  increases the
X-ray luminosity  and emission-weighted temperature by  factors of 2.9
and 1.6,  respectively, and it changes  the form of  the X-ray surface
brightness  and temperature  profiles.   At the  higher resolution,  a
simulation that  includes cooling and star formation  converts 30\% of
the cluster baryons  into stars and produces a  massive central galaxy
that substantially  alters the  cluster potential well.   This cluster
has  20\% higher  X-ray luminosity  and 30\%  higher emission-weighted
temperature   than  the  corresponding   cluster  in   the  no-cooling
simulation.  Its properties are  reasonably close to those of observed
XD clusters, with conversion of cooled gas into stars greatly reducing
the  observational  conflicts  found  by  Suginohara  \&  Ostriker  in
simulations with cooling but no star formation.  We conclude that both
resolution and  included physical processes play an  important role in
simulating the formation and evolution of galaxy clusters.  Therefore,
physical  inferences drawn  from  simulations that  do  not include  a
gaseous  component  that  can  cool  and form  stars  present  a  poor
representation of reality.
\end{abstract}


\section{Introduction}\label{introduction}
With  masses  exceeding  ${\rm  10^{13}\msun}$, clusters  of  galaxies
represent the  largest gravitationally bound objects  in the Universe.
For a number of decades, these  systems have been the focus of intense
observational  and theoretical study  in an  effort to  understand the
evolving inter-relationships between the hot diffuse intracluster gas,
the galaxies, and the inferred  dark matter component that make up the
clusters.  In recent years,  rapid advances in computer technology and
available  computational  power,   coupled  with  the  development  of
increasingly  sophisticated numerical  simulation codes,  have offered
the   possibility  of   studying  the   formation  and   evolution  of
cosmological structures like clusters in unprecedented detail.

The developmental  history of the increase in  resolution and physical
complexity  of  the numerical  simulations  has  closely followed  the
growth of computational power. Early efforts considered simple systems
consisting only of  dark matter (e.g. Peebles 1970);  the evolution of
such  systems depends  solely on  the  action of  gravity.  Since  the
number of  particles involved was relatively  small, the gravitational
forces were  calculated directly.  To  enhance the degree  of realism,
novel techniques (such as particle-mesh and tree-codes) were developed
so that  simulations with an  increasingly larger number  of particles
would be computationally feasible.

More recent simulations have also included a baryonic component.  This
was   made   possible   using   techniques  like   smoothed   particle
hydrodynamics  (SPH), to study  jointly the  distributions of  gas and
dark matter  within the cluster environment.  Unlike  the dark matter,
the evolution  of this baryonic  material depends not only  on gravity
but  also on  processes  such  as cooling  via  radiative and  Compton
processes,  heating  via  shocks  and  the  deposition  of  energy  by
supernova explosions, the  conversion of gas into stars,  etc. Even in
simulations  where  the baryonic  component  is  treated  as a  simple
adiabatic fluid, one  still needs to consider the  impact of processes
like  compressional  and shock  heating.   Such  heating modifies  the
subsequent motions of baryonic  material within the cluster potential.
The effects  of these non-gravitational processes can  be complex, and
they  can interact  in  unexpected  ways with  the  effects of  finite
numerical  resolution.  The  physics of  star formation  and supernova
feedback is poorly understood; one can introduce plausible ``recipes''
for  converting cooled  gas into  stars, but  until these  recipes are
better constrained by observations they remain a source of uncertainty
in numerical calculations.

How much do the details  of the resultant simulated clusters depend on
the   characteristics  of  the   simulation?   Specifically,   how  do
properties such  as the  mass and force  resolution and the  degree to
which  non-gravitational physical  processes are  included  affect the
results?  In  this paper we  address these questions by  considering a
family of simulations modeling  the formation of a Virgo-sized cluster
evolved  with differing resolutions  and including  different physical
processes.  A complementary study  was recently presented by Bryan and
Norman  (1998); while  they  considered a  sample  of clusters,  their
simulations were  undertaken at lower  resolution and did  not include
the cooling of gas and subsequent star formation.

In \S2 we briefly review  the results from recent simulations.  In \S3
we describe  our methods.  The three-dimensional radial  profiles of a
number of  physical properties of the simulated  clusters are compared
in \S4, more  specifically the total and dark  matter distributions in
\S4.1, the  circular velocity  in \S4.2, the  gas density  profile and
baryon fraction  in \S4.3,  the cooling timescales  in \S4.4,  the gas
temperature  distribution   in  \S4.5   and  the  entropy   in  \S4.6.
Projected,   two-dimensional   distributions,   akin   to   observable
properties  of galaxy  cluster, are  presented in  \S5. These  are the
total and dark matter distributions in \S5.1, X-ray surface brightness
in \S5.2, X-ray temperature distributions in \S5.3 and the \SZ\ effect
in  \S5.4.  We discuss  the implications  of this  study in  \S6, with
\S6.1 focusing of the effect of resolution, \S6.2 on the addition of a
hydrodynamic component  and \S6.3  the effect of  allowing the  gas to
cool and and form stars. The conclusions of our study are presented in
\S7.

\section{Past Simulations}\label{history}
Many simulations  of cluster formation  have been undertaken  over the
last  three  decades.  While  the  earliest  had  only a  dark  matter
component,  many have  included a  gas component  and, in  more recent
work,  have incorporated  the physics  of cooling  and  star formation
within the gaseous environment.  To exhaustively summarize the details
of all  the previous  simulations is beyond  the scope of  this paper.
For  comparison  purposes,  however,  we  present  the  details  of  a
representative  sample  of  recent  cluster formation  simulations  in
Table~\ref{table1}.   In  addition  to  the physical  details  of  the
simulations, such as the final cluster mass and whether or not the gas
was allowed  to cool  and form stars,  we also present  the resolution
limits  of   the  various  components.   This   includes  the  spatial
resolution, which  we define to  be twice the  gravitational softening
length, and the  dark matter and gas mass  resolution, which we define
to  be 32  times the  mass  of the  corresponding particle.   Wherever
possible  we  convert  the  gravitational softening  length  into  its
equivalent Plummer  softening value.  Most simulations  have a spatial
resolution  that is  constant in  comoving coordinates.   We  append a
``p''  to   those  that  have   a  resolution  constant   in  physical
coordinates.   For the  Eulerian  simulations we  quote  the gas  mass
resolution  at  the mean  density.   The  gas  mass resolution  for  a
virialized  object is  the number  in parentheses.   We also  define a
virial mass  limit to be  the mass of  a virialized object,  an object
with a  mean overdensity greater than  178, that has a  radius that is
twice  the  spatial resolution  scale.   This  should  be the  minimum
resolvable mass of  a collapsed object.  All values  are at $z=0$.  As
one can see from this table, the various resolutions used span several
orders of magnitude.

As  noted in  Suginohara  \& Ostriker  (1998),  these various  methods
reproduce the  generic features of  observed galaxy clusters,  such as
their  core radii,  suggesting that  the overall  picture  of clusters
forming   from  gravitationally   driven  merging   and   collapse  of
subcomponents   provides  a   reasonable  description   for  structure
formation.   While we consider  this encouraging,  we also  feel that,
owing to various numerical effects  and the exclusion of various known
physical components  and processes, these simulations  probably do not
provide  a complete  description  of the  formation  and evolution  of
galaxy clusters.

It has  been argued  that cooling should  be important only  on scales
smaller than  the cooling  radius; gas within  the cooling  radius can
cool  within a  Hubble time,  leaving gas  outside the  cooling radius
unaffected.  However,  the loss of  central pressure support  owing to
this cooling  will influence the hydrodynamic evolution  of the entire
cluster and cooling  during the early stages of  cluster formation can
remove gas from the eventual intracluster medium by converting it into
stars. For galaxy  clusters the cooling radius ranges  from 100 to 200
kpc depending on  the size of the cluster.  As  we show below, cooling
can greatly affect some physical properties of the cluster all the way
out to the virial radius.

Similarly, one  might expect the limited spatial  resolution to affect
only   scales  smaller   than  the   spatial  resolutions   listed  in
Table~\ref{table1}.  Since most of the cluster simulations resolve the
scales of observed cluster cores,  one might perhaps conclude that all
the simulations have sufficient  resolution. Although this may be true
for  the total  mass density  distribution, finite  resolution effects
modulate  the potential out  to several  times the  spatial resolution
scale.  Further, since clusters  form hierarchically, one might expect
that  it is  also necessary  to resolve  the subunits  that eventually
merge to  form the  cluster and that  not resolving the  subunits will
influence even  larger scales  (the relevant comparable  resolution is
the fraction  of the virial radius).   As we will  show, some physical
quantities  are  affected  on  larger  scales  than  those  listed  in
Table~\ref{table1}, sometimes approaching the virial radius.

Judging by Table~\ref{table1}, the  mass resolutions all seem adequate
to resolve  galaxy clusters.  But once  again, it may  be necessary to
resolve  the substructures  that merge  to form  the  galaxy clusters.
Most of the simulations appear  to have mass resolutions sufficient to
resolve several levels down  the hierarchy.  When cooling is included,
however, it may be necessary to resolve galactic scales.

\section{Methods and Models}\label{model}
For the study  presented here, most of the  simulations were performed
with TreeSPH [\cite{he89}, hereafter HK; \citet{ka96}, hereafter KWH],
a   code   that   unites   smoothed   particle   hydrodynamics   [SPH;
\cite{lucy77,gingold77}]  with   the  hierarchical  tree   method  for
computing gravitational forces  \citep{bh86,h87}.  Dark matter, stars,
and gas  are all represented  by particles; collisionless  material is
influenced  only by  gravity, while  gas is  subject  to gravitational
forces, pressure gradients, and  shocks.  Because it uses a Lagrangian
hydrodynamics  algorithm and individual  particle time  steps, TreeSPH
can  perform simulations  with the  enormous dynamic  range  needed to
study  galaxy  formation  in  a  cosmological context.   In  SPH,  gas
properties  are computed by  averaging or  ``smoothing'' over  a fixed
number of  neighboring particles; 32  in the calculations  here.  When
matter  is  distributed  homogeneously,  all  particles  have  similar
smoothing volumes.  However, smoothing  lengths in TreeSPH are allowed
to decrease in collapsing  regions, in proportion to the interparticle
separation, thus increasing the  spatial resolution in precisely those
regions where a high dynamic  range is needed.  In underdense regions,
the smoothing  lengths are larger,  but this is  physically reasonable
because  the gas  distribution  {\it is}  smoother  in these  regions,
requiring fewer  particles for an accurate  representation.  To enable
it  to  perform  cosmological  simulations TreeSPH  includes  periodic
boundary conditions,  comoving coordinates, radiative  cooling, and an
algorithm  for  star  formation   that  turns  cold,  dense  gas  into
collisionless particles  and returns supernova feedback  energy to the
surrounding medium.

To maintain accuracy  in the tree force calculation  we use an opening
angle  criterion of  of $\theta=0.7$  \citep{he89}.  We  use  a spline
kernel for  the gravitational force  softening; this has  an advantage
over  other forms  for the  softening (e.g.   Plummer);  forces become
exactly Newtonian  beyond twice the softening  length.  TreeSPH allows
particles to  have individual time  steps according to  their physical
state, so  that the pace of  the overall computation is  not driven by
the small fraction of particles requiring the smallest time steps.  To
ensure accurate integrations we use the timestep criteria described in
Quinn et al.   (1998) and KWH for each particle  with a Courant factor
of 0.3 and $\epsilon_{grav} =  0.4$.  To increase efficiency, we never
allow  the  gas   smoothing  length  to  drop  below   1/4  times  the
gravitational  softening length.   In the  hydrodynamical calculations
described  here, the  largest allowed  timestep is  $6.5  \times 10^6$
years.  The  smallest timestep was  32 times smaller, or  $2.03 \times
10^5$ years.

To  accurately  simulate the  formation  of  a  galaxy cluster  it  is
important to follow the evolution of the gravitational tidal field out
to a  large radius,  while retaining high  resolution in  regions that
eventually constitute the cluster.   The global tidal field can affect
the evolution of matter flowing into the cluster.  To accomplish these
goals using  a minimum number  of particles, hence making  the problem
computationally tractable, we use the following procedure.

First, we  used a Gelb \&  Bertschinger (1994) simulation  of a large,
uniform, periodic  volume to create  a catalog of galaxy  clusters for
further study.   The simulation tracked the evolution  of $144^3$ dark
matter particles  within a  periodic cube $100  \mpc$ (${\rm H_0  = 50
\kms{\mpc}^{-1}}$; $h  = 0.5$  throughout this paper)  on a side  in a
flat Einstein-de Sitter (${\rm \Omega=1}$) universe.  The mass of each
dark matter  particle in the simulation  is ${\rm M_{dp}  = 2.3 \times
10^{10}  \msun}$.   The  dark  matter particles  were  imprinted  with
perturbations  described  by  the  standard CDM  power  spectrum,  and
evolved  to the present  using a  ${\rm P^3M}$  algorithm \citep{ge92}
with   a  timestep  of   $1.9\times  10^7$   years  and   a  (Plummer)
gravitational  softening  length of  32.5$h^{-1}$  comoving kpc.   The
amplitude of the initial perturbations is such that the square root of
the  variance in  $8 h^{-1}  \mpc$  spheres ($\sigma_8$)  is 0.7  when
linearly extrapolated to $z=0$.  We selected one cluster, a relatively
isolated  system  with  a  circular  velocity of  about  $1000  \kms$,
representative of  a candidate  Virgo-like cluster, for  more detailed
study  employing  a  hierarchical  mass  grid  for  higher  resolution
\citep{ka93}.

The Gelb  \& Bertschinger  (1994) simulation is  used to  identify the
particles that constitute  the final cluster at the  present day.  The
new  simulation volume  is centered  on the  eventual position  of the
cluster.   The mass  resolution  in the  finest  hierarchical grid,  a
spherical  region  of  radius  $20  \mpc$  about  the  center  of  the
simulation  volume, is  the same  as that  in the  original simulation
(i.e.~${\rm   M_{dp}  =   2.3\times10^{10}\msun}$),  but   it  becomes
successively coarser at larger  distances from the cluster center.  In
this way,  only 117,000 particles  (100,000 in the central  region and
17,000  outside) are  needed  to  model the  entire  $100 \mpc$  cube,
compared to over  3,000,000 if it were modeled  at uniform resolution.
We use the same realization of the power spectrum as before and evolve
to the present  using TreeSPH (no gas component)  with a gravitational
softening  length of $35  h^{-1} {\rm\  comoving\ }  \kpc$ (equivalent
Plummer softening) and a maximum  timestep of 5.2 $\times 10^7$ years.
We use this simulation to initialize the final set of five simulations
where  the  highest  resolution  regions  are limited  to  only  those
containing particles that fall within the virial radius of the cluster
at  $z  =  0$ and  their  nearest  neighbors.   The details  of  these
simulations  are presented  below and  a summary  of  their resolution
properties is presented in Table~\ref{table2}.

The  first  of the  five  final simulations  was  a  pure dark  matter
simulation, rerun with the same mass resolution in the high resolution
regions as  in the previous  cases but with a  gravitational softening
length  of $7h^{-1}  {\rm\  comoving\ }\kpc$  (equivalent Plummer),  a
factor of five improvement  in the spatial resolution. This simulation
hereafter will be referred to as \dark.  The total number of particles
in this simulation is 68,000 with 44,000 high resolution particles and
24,000 particles of greater mass.

We perform  three further simulations  at the same mass  resolution as
\dark\ but with the addition  of a hydrodynamical gas component in the
high  resolution regions.  Of  these, two  had the  same gravitational
force resolution and  hence the same spatial resolution  as \dark\ and
the  third had a  gravitational softening  length of  $100h^{-1} {\rm\
comoving\  }\kpc$ (equivalent  Plummer).   Since we  restrict the  gas
smoothing length  to be greater  than 1/4 the  gravitational softening
length, this also has the  effect of artificially reducing the spatial
resolution of the gas.  This  low resolution SPH simulation and one of
the two high resolution  SPH simulations are evolved without supernova
heating,  radiative or Compton  cooling, or  star formation.   We will
refer to these two simulations  as \lowr\ and \adia, respectively.  By
comparing the  results of these  two simulations, we expect  to assess
the  effects of  spatial  resolution.   Note that  we  conform to  the
conventional  misuse   of  terminology  in  this   field  by  equating
``non-radiative'' gas  dynamics with ``adiabatic''  gas dynamics, even
though the gas in our adiabatic simulation is subject to shock heating
and therefore does not evolve isentropically.

In the other high  resolution hydrodynamic simulation (\clus), the gas
is allowed  to undergo cooling  via a number of  mechanisms, including
collisional   excitation  and  ionization,   recombination,  free-free
emission  and inverse  Compton cooling  off the  microwave background.
Cooling allows the  gas to collapse, where possible,  into cold, dense
knots.   In ``real''  systems, gas  in  such knots  would form  stars.
Unlike  gas, a stellar  component is  collisionless and  hence behaves
very differently  during collisions and  mergers.  Additionally, stars
can inject energy into the  surrounding medium via winds and supernova
explosions.  The fraction of the baryonic component locked up in stars
can affect both the subsequent  dynamical and thermal evolution of the
system.  Consequently,  simulations that allow the gas  to cool should
also include  a prescription for  converting gas into stars;  the {\it
Cool+SF}  simulation includes  such a  prescription for  turning cold,
dense gas  into collisionless  ``star'' particles.  The  technique and
its computational  implementation are described in detail  by KWH.  In
brief, gas  becomes ``eligible''  to form stars  if it has  a physical
density  corresponding  to  $n_{\rm   H}  >  0.1\;  \cm^{-3}$  and  an
overdensity $\rho/{\bar\rho}>56.7$  (equivalent to that  at the virial
radius  of an  isothermal  sphere).  The  gas  must also  reside in  a
convergent flow  and be locally  Jeans unstable, although  the density
criteria themselves  are usually sufficient to  ensure this.  Eligible
gas  is   converted  to  stars   at  a  rate  $d{\rm   ln}\rho_g/dt  =
-c_\star/t_g$, where  $t_g$ is the  maximum of the dynamical  time and
the cooling time.  We use  $c_\star=0.1$ for the simulations here, but
KWH show  that the  simulated galaxy population  is insensitive  to an
order-of-magnitude  change in  $c_\star$, basically  because  the star
formation rate  is forced  into approximate balance  with the  rate at
which  gas  cools  and condenses  out  of  the  hot halo.   When  star
formation occurs,  supernova heating is  added to the  surrounding gas
assuming a  standard IMF  from $0.1$ to  $100 {\rm M}_\odot$  and that
stars above $8 {\rm  M}_\odot$ become supernovae.  Each supernova adds
$10^{51}$ ergs of  thermal energy to the system.   When a gas particle
first  experiences star  formation,  a fraction  of  its mass  becomes
stellar  and  the  particle  temporarily  assumes a  dual  role  as  a
``gas-star''  particle,   with  its  contribution   to  gas  dynamical
quantities  dependent  on its  gas  mass  alone.   A gas  particle  is
converted to  a purely collisionless  star particle when its  gas mass
falls below 5\% of its initial value.  As discussed in KWH, the use of
hybrid gas-star particles allows us to avoid the computational cost of
introducing  extra  particles  into  the simulation  but  also  avoids
decreasing the  local gas resolution  in cooling regions,  which would
happen if we immediately converted gas particles into star particles.

The initial  gas particle distributions for the  three SPH simulations
are  generated by  taking the  initial particle  distribution  for the
\dark\ simulation and creating a gas particle for each high resolution
dark  particle.   The  gas  particles  are  displaced  from  the  dark
particles  by  one-half  of  a  grid  spacing in  each  of  the  three
directions.  We assume a global baryonic fraction of $\Omega_b = 0.05$
consistent   with   nucleosynthesis   values  \citep{walker91}.    The
resulting grid  has 44,000 gas particles, 44,000  high resolution dark
matter particles, and 24,000 coarser resolution particles, for a total
of 112,000 particles.

Finally, we evolve a very  high resolution dark matter only simulation
(\hire) by splitting each particle  in the \dark\ simulation that fell
within the virial radius of the cluster at $z=0$ into 27 particles and
extending  the  power  spectrum  down to  smaller  wavelengths.   This
simulation had a total of  1.3 million particles and was evolved using
PKDGRAV (Stadel  \& Quinn 1998) with a  gravitational softening length
of $0.7h^{-1}{\rm\  comoving\ }\kpc$  (equivalent Plummer) and  a dark
matter particle mass of ${\rm M_{dp} = 8.6\times10^{8}\msun}$.

All the  results presented  in this paper  are at $z=0$.   The cluster
virial  radius, defined  such that  ${\rm  \overline{\rho}(<R_{vir}) =
178\rho_{c}}$, where ${\rm \rho_{c}}$ is average density of a critical
universe, is $\sim 2\Mpc$ in  all the simulations.  The cluster virial
mass  is $4.1\times10^{14}  \msun$ and  the circular  velocity  at the
virial radius is $\sim1000\kms$.

\section{Cluster Properties: 3-D Distributions}
In this section  we examined the physical properties  of the simulated
cluster averaged  on spherical shells.   During this section  one must
remember that the cluster is neither spherical nor even homogeneous on
ellipsoidal  shells.   Interactively examining  the  structure of  the
cluster reveals sublumps, shocks and wakes whose properties are washed
out  in  the  spherical  averages.   For  example,  in  the  adiabatic
simulation there is a sublump about 1 Mpc from the cluster center that
has a  higher pressure and  density but whose temperature  is slightly
cooler  than  the surrounding  medium  ($10^7$K  vs.  a few  $10^7$K).
Trailing behind this lump is a cooler ``wake'' (few $10^6$K) extending
out to the virial radius,  which is in rough pressure equilibrium with
the surrounding  hot gas.   Ahead of  the lump is  a shock  front with
particles  as   hot  as  $10^8$K.   Structures  of   this  sort  might
conceivably  be   observable  with   the  next  generation   of  X-ray
telescopes.  Such  substructure must also  be kept in mind  during the
discussion when we refer to spherically averaged quantities.

\subsection{Total and Dark Matter Distributions}\label{total}
In Figure~\ref{total3}, we  plot the mean radial profile  of the total
(dark  $+$ baryonic)  mass density  in  the cluster,  relative to  the
universal  density, ${\rm  \rho_{c}}$.   The vertical  bars along  the
x-axis  denote the  equivalent  Plummer resolutions  derived from  the
resolution  lengths employed in  the \adia,  \dark, \clus\  and \lowr\
simulations.  The heavier bars on top of the curves indicate the radii
within which there are 32 particles; this can be considered the ``mass
resolution''  limit   of  the  simulation.   At   radii  greater  than
$\sim300\kpc$,  the cluster  mass  density profiles  in the  different
simulations are indistinguishable.  At  the virial radius they fall as
$\rho\propto  r^{-3}$.  The  effects  of resolution  and the  included
physical processes are apparent at radii less than $\sim300\kpc$.  The
mass  density profile  in the  \lowr\ simulation  becomes increasingly
shallower  with  decreasing radius,  becoming  nearly  flat  for $r  <
60\kpc$.

The radial mass profile  for the higher resolution simulation (\dark),
does  not appear to  differ significantly  from the  \hire\ simulation
except perhaps  at $r<40\kpc$, suggesting that the  total mass density
profiles  in these simulations  are free  from resolution  effects for
$r>40\kpc$.  This agrees with the conclusions of Moore et al.  (1998).
The total mass  density profiles in simulations that  include the same
physical  processes  appear to  converge  at  radii  greater than  our
defined spatial resolution limit.

The  inclusion of an  adiabatic gas  component does  not significantly
affect the total mass density  profile on large scales. Differences do
occur on small  scales, however. The profile in  the \adia\ simulation
slightly  steepens  within 60$\kpc$  and  then flattens  substantially
within 30$\kpc$, about the spatial resolution scale.  The inclusion of
cooling results  in a significant change  in the mass  profile, with a
rapid steepening within the central $\sim 100 \kpc$, approximately 1.5
times the cooling radius in the \adia\ simulation.

The mass density profile can be approximated by
\begin{equation}
\rho\left(r\right) = \frac{\rho_0}{\left(r/r_s\right)^{1.4}\left[ 
1 + \left(r/r_s\right)\right]^{1.6}}
\label{navarro}
\end{equation}
where $\rho_0 \sim  5900 \rho_{c}$ and $r_s =  370\kpc$.  This profile
has the same inner slope of ${\rm \alpha = 1.4}$ found by Moore et al.
(1998), although  the slope  in the outer  region is  slightly steeper
with ${\rm  \alpha\sim3}$ .  [More  recent high-resolution simulations
of  haloes of  different  masses  all show  an  $\sim r^{-1.5}$  inner
profile, turning  over to $r^{-3}$  at large radii~\citep{mo98,mo99}.]
Equation~\ref{navarro} matches  the density profile  of Navarro, Frenk
\& White (1997) at large radii but has a steeper central cusp.

We plot  the difference  between the profile  in equation (1)  and the
density distributions  of our simulations  in Figure~\ref{diff3}.  The
parameterization of  equation~(\ref{navarro}) accurately describes the
\hire\  simulation, and it  reasonably describes  both the  \dark\ and
\adia\  simulations except  in the  very central  regions  where these
flatten  slightly.   As  indicated  in Figure~\ref{total3},  both  the
\lowr\  and \clus\  simulations have  significant departures  from the
density profile of equation~(\ref{navarro}).

The  total  mass  density  profile  for  the  cluster  in  the  \clus\
simulation,  even though  it has  the same  spatial resolution  as the
clusters  in the  \adia\  and \dark\  simulations,  continues to  rise
steeply towards the cluster center and appears to steepen even further
within the central $20\kpc$.   The density within the central $20\kpc$
exceeds that  in the \adia\ and  \dark\ clusters by a  factor of $14$.
This behavior is caused by  additional physical processes, in the form
of cooling,  star formation and feedback, included  in this simulation
and  is caused  by baryons  collecting  in the  cluster center.   This
pooling of  baryons in the  cluster center deepens the  potential well
and modifies the dynamics in that region.

In Figure~\ref{dark3}, we show the radial density distribution for the
dark matter  component in each  of the simulations.  Since  the \dark\
and \hire\  simulations are pure  dark matter simulations,  the curves
appearing in this plot  are identical to those in Figure~\ref{total3}.
Interestingly, while  all the simulations which include  gas possess a
very  similar dark  matter  profile beyond  $\sim30\kpc$, within  this
radius it is  apparent that {\it the additional  physical processes of
gas cooling and star formation result in a radical modification in the
form of  the dark  matter profile}.   This is not  the case  for those
simulations where  the gas was  treated purely adiabatically,  and the
forms  of  the  dark  matter   profiles  for  the  \adia\  and  \lowr\
simulations are similar to those of their total mass distributions.

In each simulation with gas, the baryonic component represents a minor
fraction of  the mass at  all radii.  The  only exception is  the very
central regions in the  \clus\ simulation, where the baryonic fraction
approaches  $M_b/M \approx 10$,  revealing the  bayonic nature  of the
central regions  of this simulation.   While enhanced with  respect to
the other simulations,  the dark matter in \clus\  comprises only half
the total mass density at $20\kpc$ and even less further in.

\subsection{Circular Velocity}
The circular  velocity of a  test particle at radius  r in a spherical
mass distribution is given by
\begin{equation}
{\
V_{circ}^2 = \frac{ GM( < r ) }{r} .
}
\end{equation}
For an isothermal cluster, where  the density falls as ${\rm r^{-2}}$,
the  circular velocity is  constant with  radius.  Figure~\ref{vcirc3}
illustrates  the radial dependence  of the  circular velocity  for the
simulations presented  in this paper.   As in the previous  plots, the
vertical  bars indicate  the  two spatial  resolutions.  The  circular
velocities derived  from the various simulations are  compared to that
derived from  the profile of  Navarro, Frenk and White  (1997) defined
such that
\begin{eqnarray}
\rho(r) &\propto &\frac{1}{r}\frac{1}{(a+r)^2}  \nonumber \\
V(r)_{NFW}^2 &\propto &
\frac{1}{r}\left[\left(\frac{a}{a+r}\right) - 1 
+ \log{\left(1+\frac{r}{a}\right)}\right].
\end{eqnarray}
Such  a profile  is  over-plotted in  Figure~\ref{vcirc3}, with  ${\rm
V_{max} \sim 1125 \kms}$ at  a radius of ${\sim 500\kpc}$.  Unlike the
total mass density profiles, which are identical at radii greater than
$\sim300\kpc$, the  circular velocities for  the different simulations
only converge at ${\rm r\simgt1}\Mpc$  or slightly more than twice the
nominal spatial  resolution of the  \lowr\ simulation.  At  the virial
radius, the curves  are very similar in form  to $V_{NFW}$, decreasing
as ${\rm r^{-\frac{1}{2}}}$ with increasing radius.

For $r<1\Mpc$, the  circular velocity curves begin to  differ from one
another.  This  is very apparent  for the \lowr\ simulation;  $V_c$ is
maximum at $\sim  1\Mpc$ and drops to $\sim85\%$ of  its peak value of
$\sim$1050km/s at the spatial resolution limit.  The amplitudes of the
other simulations  all peak at  $\sim$1100km/s at 600kpc.   Apart from
slight variations in amplitude,  the remaining curves track each other
until  $\sim200\kpc$,  where  the  circular velocity  for  the  \clus\
simulation stops declining as rapidly  as the \adia, \dark\ and \hire\
simulations. The  amplitudes and slopes  of the latter three  begin to
diverge for $r\la200\kpc$.

Between $100 \kpc$\lta r \lta  $600 \kpc$ the circular velocity of the
\clus\ simulation only  decreases to $\sim95\%$ of its  peak value and
then remains essentially constant until ${\rm r\sim 20}\kpc$, where it
begins  to rise  towards the  cluster  center While  this behavior  is
expected  given the  nature of  the total  mass density  profile, this
region is  within the resolution  limit of the simulation.   Such high
circular velocities  in the  central 20 kpc  of clusters  are probably
incompatible with the observed dynamics of brightest cluster galaxies.
Conceivably better  resolution would resolve this  discrepancy, but it
may  instead  point  to  a  shortcoming in  our  input  physics.   The
difference  between  this curve  and  that  of  the \adia\  simulation
illustrates the impact  of baryon dissipation on the  structure of the
cluster potential.

In sum,  resolution affects the  circular velocity out to  typically 3
times the nominal resolution scale.  The inclusion of an adiabatic gas
component does not  have much effect on circular  velocities but a gas
component that can cool and form stars causes the circular velocity to
rise above pure dark matter calculations out to 200 kpc, about 3 times
the cooling radius of the \adia\ simulation.

\subsection{Gas Density Profile and Baryon Fraction}\label{gasdensity}
In Figure~\ref{gas3}, we  plot the radial gas density  profile for the
cluster  in  the  \lowr,  \adia,  and  \clus\  simulations.   The  two
adiabatic  simulations  approximately  converge  at  $R\sim  300\kpc$,
essentially the same as the  convergence radius for the dark matter or
total mass  density profiles and corresponding to  the nominal spatial
resolution of the  \lowr\ simulation.  The gas density  profile of the
\clus\  simulation, the  one that  includes  cooling as  well as  star
formation and its related feedback processes, does not converge to the
gas  density profiles of  the adiabatic  simulations until  the virial
radius ($\sim 2\mpc$).  We  have already commented how these processes
affect the structure of the  very central regions of the cluster, with
dissipation  producing a  strong  density peak  in  the center.   This
density peak has a strong influence on the cluster's X-ray properties,
as we discuss in \S~\ref{xray} and \S6 below.  Figure~\ref{gas3} shows
that  cooling  and  star  formation  affect the  gas  density  profile
throughout the cluster, partly by deepening the potential well, partly
by converting a significant fraction of the gas into stars, and partly
because these two effects in  turn alter the hydrodynamic evolution of
the remaining diffuse gas.

Examining the  profiles in  more detail, one  sees that all  three gas
density profiles have  a slope of $\sim2.6$ beyond  the virial radius.
At  the  virial radius  the  gas  density  profiles steepen,  possibly
indicating the  presence of a weak  shock.  At 1 Mpc  from the cluster
center, both the  \adia\ and \lowr\ gas density  profiles steepen to a
slope of  $\sim3.4$.  The steepening  is less dramatic for  the \clus\
gas density profile  which has a slope of  $\sim2.7$.  Between 300 kpc
and the virial radius $\rho_\lowr > \rho_\adia > \rho_\clus$; to first
order caused by  the response of the gas  to the different potentials.
In the  \lowr\ simulation the  central potential is shallower  and the
gas  distribution  is   more  extended.   At  300$\kpc$,  $\rho_\lowr$
flattens to  form a core  and crosses the $\rho_\adia$  profile, which
continues to  rise. At $\sim40\kpc$,  the gas density profiles  of the
\clus\ and \adia\ simulations also  cross, owing to a combination of a
rapid steepening of the \clus\ density profile and a flattening of the
\adia\  profile.  In the  \clus\ case,  gas cools  to form  a galactic
component,  much of  it found  in the  cluster center  in the  form of
stars. This effectively depletes the hot intracluster medium and forms
a giant galaxy  in the cluster center.  The  redistribution of the gas
has  important consequences for  the baryon  fraction in  the cluster.
The causes  and implications of  these processes will be  discussed in
more detail in \S6.

Figure~\ref{bar3} shows the  fraction of the mass in  baryons within a
radius,  R, normalized  by  ${\rm \Omega_b/\Omega}$,  the mean  baryon
density of the universe, which  was $0.05$ in the simulations, i.e. we
plot  $[M_b(<R)\Omega_{tot}]/  [M_{tot}(<R)\Omega_b]$.   We  plot  two
curves  for   the  \clus\  simulation;  the  upper   curve  shows  the
distribution of  the total baryonic component (in  gaseous and stellar
forms), and  the lower curve shows  the baryon fraction  based only on
the gaseous component.

The  baryon fraction  curves from  the \adia\  and  \clus\ simulations
behave differently.   As expected, both  curves asymptote to  unity as
the radius  approaches and exceeds the virial  radius.  However, going
from the virial  radius to the cluster center,  the baryon fraction in
the \adia\ simulation declines gradually until $R\sim 24\kpc$ and then
levels  off at  $\sim 20$\%  of the  universal value.   In  the \clus\
simulation, the  baryon (total)  fraction curve rises  steeply towards
the cluster center.

Looking at  the gas fraction instead  of the total  baryon fraction in
the \clus\ simulation we find a very different result.  Like the total
baryon fraction in  the \adia\ simulation, the gas  fraction in \clus\
drops gently going from the  virial radius towards the cluster center,
leveling off  at $\sim  240\kpc$.  Moreover, the  gas fraction  in the
cluster is significantly  below unity at all radii  and only climbs to
0.7 at twice  the virial radius.  The differences  between the gas and
the total  baryon fraction curves in \clus\  simulation indicates that
30\% of the baryons in the cluster are locked up in stars and that the
core of the  cluster is, in fact, a stellar  core. This is illustrated
more  graphically   in  Figure~\ref{colddense},  where   we  plot  the
distribution  of  the individual  baryonic  components  in the  \clus\
simulation; stars,  hot gas (${\rm T>10^{6.5}K}$) and  cold gas (${\rm
T<10^{6.5}K}$).  This plot is  essentially unchanged when the boundary
between  hot and  cold gas  is reduced  to ${\rm  10^{4.5}K}$.  Within
150kpc of the cluster center the baryons are dominated by stars, while
beyond  this radius,  baryons exist  mainly in  the form  of  hot gas.
Other than a small central peak and in galaxies , there is very little
cold gas in the cluster; this has all been turned into stars.

\subsection{Cooling Timescales}\label{tempsec}
In  Figure~\ref{tcool3}, we  plot the  cooling time  as a  function of
distance from the cluster center.  The cooling time is larger than the
dynamical time at  all radii in all three  simulations.  In the \lowr\
simulation none of the gas would  cool within a Hubble time if cooling
were  suddenly  included,  something   that  might  have  led  one  to
erroneously conclude that cooling  is unimportant for the intracluster
gas.  Remember  that in this and  in the \adia\  simulation cooling is
not  included during  the evolution  of  the cluster.   In the  \adia\
simulation, only within  the central 70 kpc would  the cooling time be
shorter  than the  Hubble  time  (13 Gyrs)  if  cooling were  suddenly
included.  Including  cooling in this  simulation would result  in the
eventual collapse of the gas within the central 70 kpc region.

In the  \clus\ simulation, the  only simulation that  included cooling
during the  evolution of the  cluster, the cooling radius  is actually
smaller, only 40 kpc, because the  gas density has been reduced by the
conversion of  gas into stars.  With our  star formation prescription,
the injection  of energy from supernova explosions  has no significant
effect on the thermal properties  of the intracluster gas; the cooling
time  scale in  the central  regions  is very  short and  cold gas  is
readily turned into  stars.  Any energy that is  injected into the gas
rapidly  escapes   radiatively.   Beyond  the   central  region  ${\rm
(\ga900\kpc)}$, sharp spikes are also apparent in Figure~\ref{tcool3};
these regions represents knots  of cold, collapsed gas that correspond
to galaxies within the cluster.

\subsection{Gas Temperature}\label{temp}
We  plot the radial  temperature profile  of the  gas averaged  over a
spherical  shell in Figure~\ref{temp3}.  The temperature  profiles for
all  the simulations  are similar  for radii  greater than  the virial
radius,  r \gta ${\rm 2\mpc}$, more  than 5 times  the nominal spatial
resolution  scale of  the \lowr\  simulation.  At  smaller  radii they
exhibit quite different behavior,  however.  The \lowr\ simulation has
the  shallowest rise  towards the  cluster center,  leveling off  at a
value of  $\sim$2 keV in the  central $\sim 800\kpc$.   The cluster in
this simulation has a  large isothermal core.  The central temperature
lies slightly below the virial temperature of ${\rm T_{vir} \sim 2.8}$
keV.

The  temperature profiles  of the  cluster  in the  \adia\ and  \clus\
simulations begin to differ from  one another within the central $\sim
1\mpc$.   The  temperature profile  for  the  \clus\ simulation  rises
steeply towards the cluster center where the gas temperature exceeds 6
keV.  The temperature  profile for  the \adia\  simulation  rises less
steeply, reaching a  maximum temperature of only 4  keV in the cluster
center.

The differences in  the temperature profiles of the  \lowr\ and \adia\
simulations are  caused by  resolution effects, while  the differences
between  the  the  \adia\   and  \clus\  simulation  result  from  the
additional  physical  processes  included  in the  \clus\  simulation.
Since  one  of these  processes  is  cooling,  we would  have  naively
expected the  gas in  the \clus\ simulation  to be cooler  towards the
cluster center compared to the gas in the \adia\ simulation.  However,
as  we have  already seen,  the cooling  time scale  beyond 40  kpc is
longer  than  a Hubble  time;  consequently,  cooling cannot  directly
affect the gas beyond the  very central region. The indirect effect of
cooling is to  lower the gas density outside of the  core and to raise
the central concentration of baryons, and the deepening of the cluster
potential due  to the latter has  the overall impact  of {\it raising}
the gas  temperature throughout most  of the cluster.  Within  40 kpc,
the gas  temperature does  plummet from the  peak value of  $\sim$ 6.5
keV.  We  return to the impact  of cooling on the  cluster density and
temperature profiles in \S6.

Comparing the gas temperature  profiles with the gas density profiles,
the  gas density,  in all  three simulations,  is related  to  the gas
temperature  by ${\rm  \rho_{gas} \propto  T^3}$ or  equivalently, the
pressure ${\rm P \propto \rho_{gas}^{4/3}}$  for ${\rm r > 800 \kpc}$.
In the  central regions, the relationship  approaches ${\rm \rho_{gas}
\propto   T^{3/2}}$   or   ${\rm  P\propto   \rho_{gas}^{5/3}}$,   the
relationship for an isentropic gas.  Between ${\rm 800\kpc}$ and ${\rm
40\kpc}$  the  pressure-density relationship  can  be approximated  by
${\rm P\propto\rho^{1.2}}$.  These scaling relationships break down in
the very central regions of the \clus\ cluster.

\subsection{Gas and Dark Matter Entropy}
In  the top panel of Figure~\ref{entropy3},  we  plot  radial profiles  
of  a  quantity proportional to the entropy per unit mass for an ideal 
gas,
\begin{equation}
{\rm S = \log_{10}\left({  P / \rho_{gas}^{5/3} }\right) \ ,}
\end{equation}
where  ${\rm P}={\rm  \rho_{gas} k_B  T_{gas}/\mu  m_H }$  is the  gas
pressure, ${\rm \rho_{gas}}$ is the gas density and ${\rm \mu m_H}$ is
the mean  molecular mass of  fully ionized gas (${\rm  \mu=0.6}$).  In
all three simulations, the mean entropy per unit mass of the gas at $R
> 3\mpc$   (i.e.~well    beyond   the   virial    radius)   is   quite
high---comparable  to the entropy  per unit  mass inside  the cluster.
Why does  the gas have high  entropy at such  radii?  One possibility,
suggested  by the results  of Cen  et al.~(1995)  and Cen  \& Ostriker
(1999), is  that the high entropy  of the gas outside  the cluster may
due to the  gas having been heated to  temperatures of $10^5-10^7\; K$
by shocks associated with the formation of large-scale structures such
as sheets and filaments.  An  alternative possibility is that the high
entropy of the gas outside the cluster is due to numerical effects, as
we discuss further  below.  We will address the  origin of gas entropy
in more detail in a subsequent  paper.  Here, it suffices to note that
the entropy per  unit mass inside and outside the  cluster differ by a
very small  amount.  Therefore, accretion  shocks must produce  only a
small amount  of entropy, suggesting that the  cluster accretion shock
is weak.

The gas entropy per unit  mass in all three simulations differs within
the virial  radius.  As we will  discuss in \S6, the  total entropy of
the  gas  in the  three  simulations  is  essentially the  same.   The
profiles  vary  mostly because  the  gravitational potentials  differ.
That  the  entropy  profiles   converge  only  at  the  virial  radius
highlights that differences in the potential, even if localized to the
central 300  kpc of  the cluster, can  affect the gas  distribution on
scales of the entire cluster.  The gas density profiles, as we already
discussed,  exhibit  similar  behavior.   The entropy  per  unit  mass
decreases towards  the cluster center,  dropping from $S\sim7$  at the
virial radius  to $S\sim 5.5$ at  $r=40\kpc$.  In both  the \adia\ and
the \lowr\ simulations, the intracluster gas forms a nearly isentropic
core; in the \lowr\ simulation, the radial extent of the core is ${\rm
300\kpc}$ while radial extent of  the core in the \adia\ simulation is
a factor of $\sim 3$ smaller.

In the bottom panel  of Figure~\ref{entropy3}, we plot radial profiles
of  a quantity  characterizing the  ``coarse-grain'' entropy  per unit
mass for the dark matter,
\begin{equation}
{\rm S = \log_{10}\left({  \sigma_{1D}^2 / \rho_{dark}^{2/3} }\right) \ ,}
\end{equation}
where   ${\rm   \rho_{dark}}$ is  the  local  density  of  dark matter
determined using the 64  nearest neighbours and ${\rm \sigma_{1D}}$ is
the one-dimensional velocity dispersion of those particles.

On the whole, the dark  matter entropy in all five simulations behaves
similarly in the  inner regions, declining from a  value of $\sim 5.5$
at ${\rm  r}=0.8\mpc$ to  $\sim 3.7$ at  ${\rm r=43\kpc}$.   Over this
same radial  range, the dark  matter radial density profile  scales as
${\rm \rho_{dark} \propto r^{-2}}$ for  ${\rm r > 0.1 \mpc}$ and ${\rm
\rho_{dark}\propto r^{-1.7}}$ for ${\rm r  < 0.1 \mpc}$.  If we define
${\rm     P_{dark}\equiv\rho_{dark}\sigma_{1D}^2}$,     then     ${\rm
P_{dark}\propto \rho_{dark}^{1.1}}$ for ${\rm r > 0.1 \mpc}$, which is
similar  to  the  pressure-density  relationship for  the  gas  (${\rm
P_{gas}  \propto  \rho_{gas}^{1.2}}$).   For  ${\rm r  <  0.1  \mpc}$,
however, the pressure-density relationship for the dark matter shows a
flattening trend (${\rm \rho_{dark}^{0.95}}$  whereas that for the gas
steepens to ${\rm P_{gas} \propto \rho_{gas}^{5/3}}$).  The flattening
of  the  dark  matter  pressure-density  relationship is  due  to  the
combination of  two effects:  One, the density  profile is  flatter at
${\rm r  < 0.1  \mpc}$ and two,  the azimuthally-averaged  dark matter
velocity dispersion,  which scales as  ${\rm r^{-0.1}}$ for ${\rm  r >
0.1 \mpc}$,  turns over and  decreases as ${\rm r^{0.1}}$  towards the
cluster center.

Examining the dark  matter entropy curves in detail,  we note that the
profile  for  the  \lowr\  simulation declines  somewhat  more  gently
towards  the center  than the  corresponding profiles  from  the other
simulations, a trend  that appears to be directly  correlated with the
large  force softening  length  in the  \lowr\  simulation.  The  dark
matter entropy profiles  for the \dark, \adia\ and  \clus\ also show a
tendency  to flatten on  scales comparable  to their  respective force
softening  lengths.  The  entropy profile  for the  \hire\ simulation,
which has much a smaller force softening length, shows no flattening.

At distances  greater than $\sim  1\mpc$ from the cluster  center, the
dark  matter entropy  profiles for  four of  the five  simulations are
similar.   The   only  exception  is  the  profile   from  the  \hire\
simulation.   Outside the  cluster,  the dark  matter  entropy in  the
\hire\  simulation is  markedly lower  than the  entropy in  the other
simulations.  We interpret the dark matter entropy outside the cluster
as  indicative of  particle noise  in the  simulations.   The particle
noise  is  suppressed  as   the  number  of  particles  is  increased,
explaining the difference between results for the \hire\ and the other
simulations.  We note  that since the number of  dark matter particles
and the number of gas particles in our simulations are equal, particle
noise ought to contaminate both  distributions to the same degree.  In
the  case of  the gas  particles, however,  artificial  viscosity will
convert  the random  motions due  to  noise into  thermal energy  when
streams or shells  cross, and it is possible that  the high entropy of
the  gas beyond the  cluster virial  radius in  the \adia\  and \clus\
simulations is due this type of particle noise heating.

Also, the dark matter profiles  at distances greater than $\sim 1\mpc$
have numerous  troughs.  These troughs  are due to lower  entropy dark
matter in substructures.  The  \hire\ simulation exhibits the greatest
degree of substucture and  the corresponding entropy profile is highly
serrated.   In  the  three  panels  of  Figure~\ref{darkentropy3},  we
juxtapose  the  gas  entropy  profile  and  the  dark  matter  entropy
profiles.  The offset between the  two is mostly due to the difference
in the dark matter and gas  mass density.  There is a high coincidence
between the  troughs in the gas  entropy and the  dark matter entropy,
again reinforcing the interpretation of the troughs in the dark matter
distribution  as  being due  to  the  presence  of substructure.   The
number, the  width and  the depth  of the troughs  in the  dark matter
entropy  profiles  are  directly  related  to the  resolution  of  the
simulations  and the included  physics.  In  comparing the  \adia\ and
\lowr\ simulations, there are fewer  troughs in the \lowr\ profile and
the  troughs  are shallower  and  wider  in  keeping with  the  \lowr\
simulation having fewer, more ``fluffy'' clumps.

Comparing  the behaviour of  the dark  matter and  gas entropy  in the
three simulations  with gas, we find  that, except for  an offset, the
two track  each other very  well on scales  ${\rm r >  0.2\mpc}$.  The
offset between the two is due to the fact that the dark matter density
is larger  than that  of the  gas and the  dark matter  temperature is
typically  lower than  the gas  temperature by  a factor  of  0.8.  On
smaller scales, however, the  dark matter entropy profile continues to
decrease towards the cluster center  whereas the gas entropy curve, at
least in  the \adia\  and the \lowr\  simulations, flattens to  form a
nearly  isentropic  core.  (The  gas  entropy profile  in  the  \clus\
simulation also shows signs of flattening in the inner ${\rm 0.1\mpc}$
but the steep decline in the gas entropy at the very center due to the
cooling of the  gas makes the comparison with  the dark matter entropy
less straightforward.)  While this difference could be partly caused by
caused by  differences in numerical  treatment of SPH and  dark matter
particles, we speculate  that it is primarily a  result of differences
in the physical properties of  the gas and dark matter components.  In
the case of the gas, short-range forces prevent colliding gas elements
from interpenetrating,  resulting in  the formation of  shocks.  These
shocks  convert the  kinetic energy  of the  gas into  thermal energy;
depending on the geometry of  the shocks, the extent of acoustic waves
generated by  the shocks and  the details of the  artificial viscosity
used  in  the simulations,  the  energy  can  be distributed  over  an
extended region.  Dark  matter, on the other hand,  is a collisionless
fluid and  will interpenetrate.   This interpenetration can  result in
dark matter elements with very different velocities occupying the same
small but finite spatial volume,  resulting in a ``mixed'' system. The
coarse-grain entropy, as we have defined  it, is a measure of how well
the system is  mixed in this sense.  The  dark matter entropy profiles
of our simulated clusters indicate that the dark matter in the central
regions is  not as well  mixed as in  the outer regions. This  is also
borne out  by the  decline in the  dark matter velocity  dispersion on
scales smaller than ${\rm 0.2\mpc}$.   The extent of mixing depends on
the initial state  of the system as well as  on constraints imposed by
integrals  of  motion  (Tremaine,  Henon  and  Lynden-Bell  1986)  and
therefore, is likely to be affected by the physical characteristics of
the  system, such  as the  angular momentum  distribution of  the dark
particles, as well as by numerical effects, such as particle and force
resolutions.

\newcommand{\x}{{\bf X}} 
\newcommand{\y}{{\bf Y}} 
\newcommand{\z}{{\bf Z}}
\newcommand{\X}{{\bf X}} 
\newcommand{\Y}{{\bf Y}} 
\newcommand{\Z}{{\bf Z}}

\section{Cluster Properties: Projected Distributions}
Simulations of galaxy  clusters have a wealth of  structure on a range
of  scales.    Observationally,  however,  we  are   restricted  to  a
two-dimensional view of these systems, in the form of X-ray luminosity
and temperature distributions, and  maps of the projected surface mass
density derived  from the analysis of  gravitational lensing features.
In  the  following  section  we  present the  radial  distribution  of
projected quantities.   For this, we consider  a box with  a length of
${\rm  4\Mpc}$ on a  side, twice  the virial  radius, centered  on the
deepest  point in  the cluster  potential.  All  quantities  are first
smoothed in three dimensions with an adaptive SPH-like kernel over the
32 nearest  neighbors.  This  smoothed distribution is  then projected
along the required axis [see \citet{ka93}].

\subsection{Total and Dark Matter Surface Mass 
Distributions}\label{proj_all}     The     left-hand     column     of
Figure~\ref{mass2}  plots the  radial  profiles of  the the  projected
total mass density and the  right-hand column plots the radial surface
mass density profiles  for only the dark matter.   Each panel, labeled
\x, \y\  and \z, plots projections  along one of  the simulation axes;
the total mass within the box is $1.3\times10^{15}\msun$.

As expected from our analyses  of the three-dimensional total and dark
matter distributions in the different simulations, the projected total
and  dark matter mass  distributions in  the \lowr\  simulation appear
similar.  At  large radii, the  distributions fall off  as $R^{-1.8}$,
which   agrees  with   what  one   would  have   expected   given  the
three-dimensional  distributions.   Towards  the cluster  center,  the
profiles  flatten,  eventually forming  a  core,  which  again is  not
surprising  given  the   nature  of  the  three-dimensional  profiles.
However, the  size of the core  depends on the viewing  angle.  In the
\x\ and  \y\ projections,  the core begins  to form  at $\sim300\kpc$,
consistent   with   what   one   would   have   expected   given   the
three-dimensional  profiles, but  in the  \z\ projection  the  core is
smaller  by a  factor  of $\sim  3$  and the  central surface  density
exceeds that in the \x\ projection  by a factor of $\sim 1.6$ and that
in the \y\ projection by a factor of $\sim 2.6$.  It appears that this
enhancement is  not due  to the fortuitous  presence of a  subclump of
matter  along the  \z\ axis  but rather  it is  caused by  the central
regions  of the  cluster being  ellipsoidal rather  than  spherical in
shape with the major axis aligned along the \z\ axis.

The total and dark matter surface mass density profiles for the \adia\
and \dark\ simulations  are also very similar.  There  is a dependence
of  the core  size  and central  peak  projected mass  density on  the
orientation of the cluster as  seen in the \lowr\ simulation, although
neither  flatten enough  to make  a core  in the  central  region. The
projected  mass  density in  the  \adia\  simulation  has a  slope  of
$\sim0.63$ at radii  less than 100 kpc in the  \x\ and \y\ projections
while  the projected  mass  density  in the  \dark\  simulation has  a
slightly shallower  slope of $\sim0.5$.  As in  the \lowr\ simulation,
the \z\  projection exhibits a  markedly different behavior,  with the
\adia\ simulation having a steeper  slope of $\sim0.88$ in the central
regions and the \dark\ simulation having a slope of 0.66, both steeper
than the slopes in the \x\ and \y\ projections.  The central projected
mass density in  the \z\ projection of the \adia\  exceeds that of the
other orientations by a factor of 1.7, while that of \dark\ simulation
is 1.6 times higher.

The projected surface mass density profile in the \hire\ simulation is
virtually indistinguishable from its lower resolution counterpart, the
\dark\  simulation.    The  projected  dark  matter   and  total  mass
distributions   in  the   \clus\   simulation  are   similar  to   the
corresponding projections of the  \adia\ and \dark\ simulations for $R
\simgt  50\kpc$.  At smaller  radii, however,  the total  surface mass
density exhibits  a sharp up-turn,  with a slope of  $\sim1.15$.  This
peak is not as prominent in  the dark matter projections and is caused
by  the large central  galaxy.  Examining  the profiles  more closely,
while both the  total and dark matter surface  density profiles in the
\x\ and  \y\ projections are indistinguishable  from the corresponding
\adia\  and  \dark\  profiles  beyond  the central  peak  region,  the
projected mass  density profile  of the \clus\  simulation in  the \z\
projection  lies below  that  of  the \adia\  and  \dark\ profiles  at
clustercentric  distances up to  $\sim200\kpc$.  As  in all  the other
simulations,  the central  surface density  in the  \z\  projection is
greater  than that  in the  other  two projections,  although the  \z\
excess is only a factor of 1.25 in this case.

Even though  the mass  within the  virial radius is  the same  in each
simulation, each exhibits different lensing signatures. It is possible
for the \dark\ cluster  to be super-critical to gravitational lensing,
producing  giant arcs,  while  the \lowr\  simulation  lies below  the
critical mass threshold and  produces no strong lensing features.  The
gravitational lensing characteristics of a galaxy cluster, such as the
radial  positions and  widths of  giant arcs  and the  distribution of
weakly lensed images, depend on  the core radius and central value and
slope of  the projected  mass distribution (e.g.~Hammer 1991),  not on
the total cluster mass. A  small, but very centrally condensed, galaxy
cluster can produce a much stronger lensing signal than a more massive
cluster that is very diffuse.  Using collisionless N-body simulations,
Bartelmann et  al. (1998) argue  that the observed incidence  of large
arcs in clusters rules out most  of the parameter space of CDM models,
leaving only those models with low $\Omega_{\rm matter}$ {\it and} low
$\Omega_{\Lambda}$.  They note  that the effects of cD  galaxies are a
source  of uncertainty in  their conclusion,  and our  results suggest
that the effects of radiative  cooling and star formation could easily
move some  cosmological models  from the ``unacceptable''  category to
the ``acceptable'' category.

Due to  differences in the projected mass  distributions the simulated
galaxy clusters have  a lensing signature that depends  on the viewing
angle.  Therefore, the mass  distribution inferred from an analysis of
the  gravitational lensing  characteristics would  also depend  on the
viewing angle.  The  X-ray profile, which we discuss  in detail in the
next subsection, does not depend strongly on the viewing angle and has
about  the same  profile  when  viewed from  all  angles.  This  might
partially  explain  the observed  discrepancy  between cluster  masses
derived  from gravitational  lensing techniques  versus  those derived
from  X-ray analyses  (e.g.~Allen  1998) and  will  be addressed  more
carefully in a forthcoming paper.

\subsection{X-Ray Surface Brightness}\label{xray}
As  illustrated  in Figure~\ref{temp3},  the  gas  that permeates  the
galaxy cluster has  a temperature of several keV,  and such an ionized
medium emits  copious amounts of  X-rays.  Over the last  two decades,
observations of galaxy clusters with space-borne X-ray telescopes have
provided important insights on the physical properties of intracluster
gas.  With  the assumption of spherical symmetry  and isothermality, a
``fundamental  plane''-like  relation  between the  X-ray  luminosity,
temperature  and mass  (e.g.~Edge \&  Stewart 1991)  has been  used to
probe the cluster potential, making  X-ray analysis one of the primary
tools for investigating clusters.

The  inclusion of  a baryonic  component into  the  simulated clusters
allows  us  to  similarly   study  such  X-ray  characteristics.   The
temperature and  density of the gas  is used, via a  Raymond and Smith
plasma  code \citep{ra77},  to  determine the  X-ray  emission at  any
location  in  the  cluster.   We  project this  emission  along  three
orthogonal lines  of sight, as we  did with the  mass distribution, to
produce two-dimensional  X-ray surface  brightness maps using  the SPH
spline kernel  smoothing [see  Tsai et al.   (1994) for  details].  We
assume  the  metal  abundance  of  the  gas  is  0.3  Solar---a  value
considered typical for galaxy clusters \citep{mu97}.

We plot  the X-ray surface brightness distributions  for the simulated
clusters in Figure~\ref{xray2}.  The left-hand panels plot the surface
brightness  distributions  for   a  ${2\rightarrow10}$  keV  bandpass,
similar  to the  ASCA Gas  Imaging Spectrometers  (GIS). We  have not,
however, filtered  the X-ray surface brightness  with any instrumental
response functions and one should  consider this to be a ``raw X-ray''
view of the cluster in this bandpass. This also allows a comparison to
the  X-ray  luminosities  in  recently  tabulated  surveys  of  galaxy
clusters [e.g.~David  et al. (1993)].  The right-hand  panels plot the
X-ray surface  brightness distributions for  a $0.1\rightarrow2.4$ keV
bandpass  similar  to the  ROSAT  X-ray  telescope  filtered with  the
response function  of the High  Resolution Imager (HRI); these  can be
considered a true X-ray telescope view of the simulated cluster.

For \clus, \adia\ and \lowr\  simulations, the X-ray luminosity of the
clusters in the ${\rm  2\rightarrow10}$keV bandpass are 3.14, 2.58 and
0.88${\rm \times  10^{43}\ erg\ s^{-1}}$ respectively  (these are also
summarized in  Table~\ref{table3}).  In the  same band pass  the Virgo
cluster   has  an   observed   X-ray  luminosity   of   ${\rm  L_x   =
1.16\times10^{43}\  erg\  s^{-1}}$, with  a  bolometric luminosity  of
${\rm L_x^{Bol} =  2.66\times10^{43}\ erg\ s^{-1}}$ \citep{da93}.  The
X-ray flux in  this same bandpass is within a factor  of three of this
observed value for all the simulated clusters.

As with  the total X-ray  flux, the X-ray surface  brightness profiles
depend on the details of the simulation, but for each simulation these
profiles  have  generic features  that  are  independent  of both  the
orientation and  the energy band  under consideration.  Beyond  1 Mpc,
about  half  the  virial  radius,  all the  X-ray  surface  brightness
profiles  are similar.   Within  the inner  $\sim300\kpc$, the  \lowr\
simulation profile rapidly turns over, forming a flat X-ray core.  The
X-ray surface  brightness in the  \adia\ simulation continues  to rise
all the  way into the center  of the cluster, slowly  turning over but
never fully  flattening to form  an apparent core.  The  X-ray surface
brightness profile of the \clus\ simulation falls below the others for
$R\simlt1\mpc$.   Then, rising  slowly, it  crosses the  X-ray surface
brightness profile  of the \lowr\ simulation inside  its core.  Within
the  central  $50\kpc$,  about   the  cooling  radius  in  the  \adia\
simulation, the  profile of the \clus\ simulation  rapidly rises above
the  profile of  the \adia\  simulation.   The cluster  in the  \clus\
simulation has a  bright X-ray core, a factor of  10 brighter than the
central value in the \adia\ simulation.

The  \clus\  X-ray  surface  brightness  profiles  have  a  number  of
additional  X-ray  ``spikes''  caused  by emission  from  cooling  gas
associated with galaxies.  Such systems  cannot form in the \adia\ and
\lowr\  simulations.    However,  superimposed  upon   their  smoother
profiles,  one can  identify additional  bumps.  These  are associated
with a prominent peak in the \clus\ distribution. These deviations are
caused by  the interaction of  the intracluster medium with  a merging
subclump.

In contrast to the projected  mass density profiles, the viewing angle
has  very little effect  on the  shape or  normalization of  the X-ray
profiles.  The only  exception is the location of  the galaxy peaks in
the \clus\  profiles.  The slopes of  the X-ray profiles  do depend on
the energy band, however.

To  summarize, resolution  affects X-ray  surface  brightness profiles
within the nominal spatial resolution scale, flattening them to form a
core.  The  addition of cooling  and star formation affects  the X-ray
surface brightness  profiles out to  one half the virial  radius, both
reducing their  brightness and changing their shape  to an approximate
power law. Within  the cooling radius, the profiles  in the simulation
with cooling and  star formation rapidly rise, forming  a bright X-ray
core.  Such sharp central peaks are observed in a substantial fraction
of  galaxy   clusters  and   are  interpreted  as   the  observational
consequence of  a cluster ``cooling  flow'' (see, e.g.,  Fabian 1994);
whether or  not this  is the  case in our  simulated clusters  will be
addressed in \S6.

The X-ray surface brightness distributions of observed galaxy clusters
are often fit by a simple parameterized model given by
\begin{equation}
{\rm
I_x \propto \left[ 1 + \left( \frac{r}{r_c} \right)^2 \right]^{
\frac{1}{2} - 3\beta} } ,
\label{beta}
\end{equation}
where ${\rm r_c}$ is the core radius and $\beta$ controls the slope of
the profile. Observed clusters have  core radii of $100-600\ \kpc$ and
$\beta   \sim  0.7$  \citep{jo84}.    We  fit   this   model  to   the
2$\rightarrow$10 keV X-ray profiles in Figure~\ref{xray2}.  This model
fits  the \lowr\  simulation  X-ray  profiles with  a  core radius  of
560$\kpc$ and $\beta=1.4$.

Figure~\ref{beta2}  shows  fits  of  the $\beta$-model  to  the  three
projected X-ray  surface brightness profiles of  the \clus\ simulation
(left-hand panel)  and the \adia\ simulation  (right-hand panel).  The
presence of the central X-ray cusp (or bright galaxy) and the galactic
spikes  in the  \clus\ simulation  complicates the  fitting procedure.
Following the analyses of X-ray emissions from XD clusters by Jones \&
Forman (1984), we excise the  central cusp region.  We also excise the
galactic spikes.   We use only  those locations marked with  a circled
point  in  the fits.   We  fit  two  different $\beta$-models  to  the
simulations:  one uses the  full range  of the  X-ray profile  and the
second  uses  only  a  restricted  range,  from 50  kpc  to  750  kpc,
approximately the  range used for observed  clusters \citep{pe98}.  We
superimpose these  fits over  the profiles in  Figure~\ref{beta2} with
the dot-dashed line  corresponding to the fit over  the full range and
the  solid   line  to  the   fit  over  the  restricted   range.   The
$\beta$-models provide a reasonable fit to the data, with the fit over
the  restricted   range  accurately  reproducing   the  X-ray  surface
brightness  within   0.5Mpc  (neglecting  the  central   cusp  in  the
\clus). The result  of fitting to the restricted  data range decreases
the  core  radius and  causes  the  slope  to become  shallower,  i.e.
$\beta$  changes from  $\sim1$  to $\sim  0.7$.  Observationally,  the
fitting of X-ray profiles find a $\beta\sim0.7$, while measurements of
$\beta$ considering  the gas temperature and cluster  dynamics yield a
value  of $\sim1$  [e.g.~Jones \&  Forman (1999)],  a  situation often
referred to  as the  $\beta$-discrepancy.  While several  solutions to
this problem have been put forward \citep{ed91,ba94,ca96}, the results
found with our numerical simulation  suggest that the situation may be
resolved  if  X-ray  profiles  could  be determined  to  larger  radii
[e.g. Navarro,  Frenk \& White  (1995)].  The addition of  cooling and
star formation  does not  greatly affect the  $\beta$-model parameters
when the  central region is  excluded.  In fact, the  model parameters
are much more  sensitive to the range over which  the fit is performed
than to the inclusion of cooling and star formation.

\subsection{X-Ray Temperature}
In  addition to  measuring the  X-ray brightness,  recent  advances in
X-ray detector technology have  provided satellite telescopes, such as
ASCA,  with increased  spatial and  spectral resolution,  allowing the
determination of X-ray emission  weighted temperature maps of clusters
\citep{ma98}.  Such  observations provide a  more direct probe  of the
physical  properties  of  the  intracluster gas,  revealing  that  the
majority of clusters deviate substantially from isothermality and have
strong temperature gradients,  with many being a factor  of two hotter
in the central regions than at the virial radius.

Since the X-ray emissivity as  a function of position in the simulated
cluster is  known, it is straightforward to  use the three-dimensional
temperature information  to determine the  projected X-ray temperature
profiles, weighted with the X-ray emission in the chosen bandpass, for
the  simulated  clusters.   We  plot  these  temperature  profiles  in
Figure~\ref{temp2}.   The temperature  profiles depend  on  the energy
band  because the  emissivity in  each band  samples gas  at different
temperatures.

In  the $2\rightarrow10$ keV  band the  \lowr\ simulation  exhibits an
almost   flat,  isothermal-like   profile.    The  emission   weighted
temperature  of  $\sim  2$  keV  is slightly  lower  than  the  virial
temperature of ${\rm T_{vir}\sim2.8}$  keV.  In the outer parts, ${\rm
r  > 1\Mpc}$,  the \clus\  and \adia\  temperature  profiles gradually
decline, falling by a factor of  $\sim2$ between 1 and 4Mpc, and hence
deviate from  an isothermal temperature profile.  Within  1 Mpc, about
half  the virial  radius,  both temperature  profiles  begin to  rise,
peaking at  ${\rm \sim2T_{vir}}$ for  the \clus\ simulation  and ${\rm
\sim1.4T_{vir}}$  for the \adia\  simulation.  The  \clus\ temperature
profile falls  to ${\rm \sim1.8T_{vir}}$  in the very  central regions
due to  the presence of a small  amount of cold gas.   The profiles in
the \adia\ and \clus\ simulations are similar to the profiles observed
in galaxy  clusters \citep{ma98}.  The simulations  have mean emission
weighted  temperatures  of  ${\rm  \sim 1.2T_{vir}}$  for  the  \adia\
simulation  and   ${\rm  \sim  1.6T_{vir}}$   for  \clus\  simulation.
However,  in  contrast   to  the  spherically  averaged  3-dimensional
temperature  profiles shown  in Figure~\ref{temp3},  projection causes
the X-ray emission  weighted temperature to have a  less dramatic rise
towards  the center  of the  cluster.  Also,  the resolution  of X-ray
telescopes tends to  be rather poor and much of  the structure seen in
Figure~\ref{temp2} would  not be resolved, resulting  in an apparently
shallower gradient in the observed temperature distribution.

In  the  ROSAT pass-band,  the  \lowr\  simulation  no longer  has  an
isothermal temperature profile but  slowly rises to ${\rm 0.7T_{vir}}$
in  the cluster  center.  Similarly,  the \adia\  simulation  rises to
${\rm  1.4T_{vir}}$,  while  the  \clus\  simulation  rises  to  ${\rm
2T_{vir}}$.  In the very center, the temperature profile in the \clus\
simulation  drops sharply  to  ${\rm 1.4T_{vir}}$,  again  due to  the
presence of  cold gas.  This decline  is more pronounced  in the ROSAT
band, since it is more sensitive to emission from colder gas (remember
that the X-ray spectrum has been filtered with the instrument response
of the ROSAT/HRI camera).   At ${\rm r\simgt1.4\Mpc}$, both the \clus\
and \adia\  simulations follow the  temperature profile of  the \lowr\
simulation, but within 1 Mpc both depart significantly, rising to peak
temperature of ${\rm \sim1.4T_{vir}}$.

Finite  numerical  resolution  affects  the projected  X-ray  emission
weighted gas  temperature profile  even on scales  that are  more than
$\sim4$  times the nominal  spatial resolution  scale.  In  the \lowr\
simulation  this  extends  all  the  way out  to  the  virial  radius.
Improving the spatial resolution by  a factor of $\sim 6$ increases the
mean  emission weighted  temperature of  the cluster  by  almost 60\%.
Adding cooling  and star formation  increases the mean  temperature by
another  30\%, to  more than  1.5  times the  virial temperature,  and
increases the  central temperature by  almost 50\% over that  found in
the \adia\ simulation.

\subsection{\SZ\ Decrement}
The bulk and thermal motions of  the hot gas that pervades the cluster
environment can affect the spectrum of the Cosmic Microwave Background
(CMB)  via Compton  scattering.   This induces  an  anisotropy in  the
observed   CMB  distribution,   measured   as  a   decrement  in   the
Rayleigh-Jeans temperature  of CMB radiation  in the direction  of the
cluster  \citep{su72}.  Such  \SZ\  decrements have  been observed  in
several clusters at radio wavelengths [e.g. Myers et al. (1997)] , and
more recently  at sub-mm wavelengths \citep{la98}, and  are proving to
be  a useful probe  of the  intracluster medium,  as well  as offering
clues to the values of cosmological parameters \citep{fu98}.

The inclusion of a baryonic  component in the simulations allows us to
investigate   the   \SZ\   decrement   in  the   synthetic   clusters.
Figure~\ref{sz2} plots the combined  local kinematics and thermal \SZ\
decrement, $\Delta T/T$, as a function  of radius (note that we do not
consider the kinematic  effect due to the bulk  cluster motion).  This
quantity is  independent of frequency in the  Rayleigh-Jeans region of
the  spectrum.   Each  projection  axis   is  the  same  as  those  in
Figure~\ref{mass2}.  Only in the  very outermost parts of the cluster,
beyond the  virial radius,  does the decrement  depend on  the viewing
angle, although even  there the differences, and the  \SZ\ signal, are
small.   In  the  central  regions,  ${R<400  \kpc}$,  the  individual
profiles diverge,  with the \lowr\ simulation turning  over.  Both the
\clus\ and  \adia\ simulations have decrements that  continue to rise.
Although the  \clus\ simulation appears shallower at  larger radii, in
the very center it peaks, with  a central value similar to that of the
more gently  rising \adia\ simulation.   These features do  not depend
strongly on  the cluster  orientation. In the  central regions  of the
cluster, the thermal component  dominates the kinematic \SZ\ effect by
an  order  of  magntude.   This  is consistent  with  the  small  flow
velocities $(\sim50\kms)$ seen in this region.

\section{Discussion}
It  is  apparent  from  the  previous  Sections  that  both  numerical
resolution and input  physics can affect the evolution  of a simulated
cluster. Here, we consider the impact of these individually.

\subsection{Resolution}\label{rescon}
Examining  Figures~\ref{total3}  and~\ref{dark3}   one  can  see  that
changing the force  resolution affects the total mass  and dark matter
profiles only  within about twice  the spatial resolution  scale.  The
\lowr\  and \adia\  simulations  have identical  profiles beyond  that
point.  The mass profiles of  the \lowr\ simulation show a significant
flattening that  begins on scales  of twice the force  resolution. The
\adia\  mass profiles continue  to rise  but also  eventually flatten,
again  on a  scale about  twice  the spatial  resolution.  The  \dark\
simulation,  consisting of  only  dark matter,  has  the same  spatial
resolution  as   the  \adia\  simulation.   Compared   to  the  \hire\
simulation, which  is also a pure  dark matter simulation,  but with a
spatial resolution ten times  smaller than the \dark\ simulation, once
again both the total mass  and dark matter profiles of each simulation
are the  same until approximately twice the  \dark\ simulation spatial
resolution scale,  where the  \dark\ simulation mass  profiles flatten
slightly and  depart from the \hire\ simulation  mass profiles.  Moore
et al. (1998) also find a  continual steepening of the mass profile as
the  spatial  resolution  is  increased  in  their  dark  matter  only
simulations, but the mass  profiles eventually converge as the spatial
resolution  scale  became small  enough  (at  fixed mass  resolution).
However, they also  find that the mass profile  becomes steeper as the
mass  resolution  is  increased,   which  accounts  for  some  of  the
differences in the mass profiles of the \dark\ and \hire\ simulations.

The  degree to which  particles can  bind together  gravitationally is
limited by  the hardness  of gravitational interactions,  resulting in
deeper potentials  for smaller  softenings.  But mass  resolution also
plays a  part: the smaller the  particle mass, the  denser the central
core  can   become,  steepening   the  potential.   As   mentioned  in
Section~\ref{proj_all}, the  steepness of the central  core can affect
the  gravitational lensing  characteristics of  a cluster  of galaxies
[e.g.~Bartelmann  (1996)]  and one  might  draw erroneous  conclusions
about the expected lensing signature  from a model cluster if the mass
profile has not converged.

While  resolution affects  the form  of  the mass  density profile  on
scales of  twice the spatial resolution, the  cluster potential, which
depends on  the integrated  mass profile, is  modified on  much larger
scales.  Therefore  properties that probe  the potential, such  as the
circular velocity and the  temperature, are affected more globally. In
Figure~\ref{vcirc3}, the difference  between the central mass profiles
of the various simulations leads to different circular velocity curves
on scales approaching half the virial radius.  The distribution of gas
within the cluster environment also depends on the form of the cluster
potential.  In Figure~\ref{temp3},  the shallower cluster potential in
the \lowr\  simulation results in less compressional  heating than the
higher resolution  simulation whose  potential is deeper,  causing the
temperature  to be  affected out  to the  virial radius.  Finally, the
total X-ray luminosity of the  cluster also depends on the resolution.
The higher resolution \adia\ cluster  has, as expected, a higher X-ray
luminosity than the \lowr\ cluster.

\subsection{The Addition of Baryons}
Before discussing  the detailed evolution of the  baryonic material in
these simulations, it is prudent to investigate whether or not the gas
is  artificially heated  by numerical  effects,  specifically two-body
interactions similar  to those responsible for  heating and subsequent
core  collapse  in  stellar   systems.   Such  effects  were  recently
investigated  by Steinmetz  \& White  (1997), whose  analysis revealed
that such heating effects  can significantly modify the properties and
hence  evolution of  the  gaseous component.   The  magnitude of  this
effect  depends mainly  on  the  mass of  the  dark matter  particles,
although the  relative gas particle  to dark matter particle  mass can
mildly  influence the  cluster evolution.   Steinmetz \&  White (1997)
demonstrate that if  the mass of the individual  dark matter particles
is  below that of  a critical  mass, given  by ${\rm  M_{crit} \approx
5\times10^9 \sqrt{T_6} M_\odot}$, where ${\rm T_6}$ is the temperature
in units  of ${\rm  10^6K}$, then spurious  heating affects  should be
unimportant.  In the high resolution regions of our simulations, where
the temperature  is several keV,  ${\rm M_{crit} \sim  4\times 10^{10}
M_\odot}$; the  masses employed in  the simulation are ${\rm  M_{gp} =
1.2\times10^{9} \msun}$  and ${\rm M_{dp}  = 2.2\times10^{10} \msun}$,
respectively.  We conclude, therefore,  that our simulations should be
free from spurious two-body gas-dark matter heating effects.

In comparison to the \dark\ simulation, including a baryonic component
in the  form of an adiabatic  gas (\adia) does not  greatly affect the
overall  properties of the  cluster.  Looking  at Figures~\ref{total3}
and~\ref{dark3}, the total mass and the dark matter profiles both have
a  very similar  form outside  of  a radius  a few  times the  spatial
resolution.   Within this distance,  the mass  profiles of  the \adia\
simulation  rise slightly  more  steeply than  the \dark\  simulation,
before  flattening. The  \dark\  profile continues  to  rise into  the
center  of the  cluster.   As  with the  effects  of resolution,  this
redistribution of matter in the cluster, and hence modification of the
cluster potential,  influences the evolution of the  cluster on scales
several times the spatial resolution.

\subsection{The Addition of Cooling and Star Formation}
The inclusion of additional physical  processes in the form of cooling
and star  formation can radically affect the  evolution and subsequent
properties of  a cluster.  For  example, consider the profiles  of the
total  mass   and  dark  matter   distributions  (Figures~\ref{total3}
and~\ref{dark3}). The \adia\ simulation  flattens as it approaches the
center, but the \clus\ simulation  continues to rise rapidly and has a
central peak. Figure~\ref{colddense} indicates that this baryonic core
is mainly stellar.

How did the  baryons pool into the center?   Two possibilities present
themselves, both of  which may be occurring in  the simulated cluster.
First, the  dense gas at the  center of the cluster  can dissipate its
energy by radiative  cooling, collapse to very high  density, and form
stars  -- in  essence, a  ``cooling flow''  creates a  massive central
galaxy.  Second, dynamical friction and stripping of subclumps as they
plunge into the cluster center can deposit {\it stellar} material into
the cluster  core \citep{ba93}.  Before the  cluster itself assembles,
cooling  allows gas  to  condense into  the  centers of  ``subunits,''
individual galaxies  or galaxy  groups. If the  gas is cold  and dense
enough, it turns into stars.  As a subunit merges into the cluster its
less dense outer  parts, comprised mostly of dark  matter, are tidally
stripped first  and deposited further  out in the  cluster.  Dynamical
friction drags  the dense  core of the  subunit closer to  the cluster
center, where eventually  the tidal field is strong  enough to disrupt
it, depositing this baryon rich material deep in the cluster potential
well.   This process  can  cause segregation  of  baryons towards  the
center of  the cluster even  if the cooling  times in the  cluster are
long.

To fully differentiate between  these possibilities one needs to trace
the dynamical evolution of  the baryonic material during the formation
and subsequent evolution  of the cluster. This is  beyond the scope of
this  paper   which  focuses  upon   the  end-point  of   the  cluster
simulations, but will be the subject of a forthcoming article in which
the cluster  properties will be traced through  the earlier simulation
outputs. We can,  however, address the question of  whether or not the
\clus\ simulation has a cooling flow at the present time.  We plot the
gas  mass accretion rate  through spherical  shells into  the cluster,
defined to be $\dot{M}=  4\pi \rho_{gas}(r) r^2 V_r(r)$, normalized by
the rate required for the cluster to accumulate its mass of gas within
the  virial radius  (e.g. ${\rm  \Omega_B M_{vir}}$)  steadily  over a
Hubble   time   (i.e.    $\dot{M}\approx  1\times10^{3}\msun/$yr)   in
Figure~\ref{mass_accretion3}.   Each of  the  simulations has  similar
rates of accretion, with both the \clus\ and \adia\ simulations having
essentially  identical profiles  into $40  \kpc$.  Since  there  is no
cooling in the \adia\ simulation, this implies that presently there is
{\it no} significant cooling flow  in the \clus\ simulation on a scale
${\rm \ga40\kpc}$.  Considering  Figure~\ref{xray2}, it is within this
radius that  the X-ray profiles  of the \clus\ simulations  possess an
X-ray cusp;  when coupled with the  increase in the  mass accretion in
this  region,  the  evidence  points  towards  the  \clus\  simulation
possessing a  cooling flow  within $40 \kpc$.   This scale  is smaller
than  typical scale  of the  cooling  region [${\rm  150\kpc}$ of  the
cooling flow  clusters \cite{pe98}].  This cooling  flow, however, has
no  significant  impact  on  the  inflow of  mass  on  larger  scales,
suggesting that the latter is the result of recently accreted material
being brought to a halt and redistributed with the cluster.

We also  examine the radial  infall velocity of  the gas in  the three
simulations.   In  Figure~\ref{dynamic3}  we  plot the  ratio  of  the
spherically averaged radial infall velocity to the dynamical velocity,
defined    as   ${\rm    r/t_{dyn}}$    where   ${\rm    t_{dyn}\equiv
(G\bar\rho_{tot})^{-1/2}}$.   The  mean   radial  infall  velocity  is
approximately $350\kms$ at distance of $4\mpc$ from the cluster center
and  scales roughly as  $\left|V_r\right| \propto  r$ from  $4\mpc$ to
$\sim  0.5\mpc$  {\it  in  all  three  simulations}.   Therefore,  the
characteristic time  scale for the  radially infalling flow  over this
range in radii is nearly constant at ${\rm t_{flow}}\sim 10\Gyr$.  For
$r >  3\mpc$, the flow  time is equal  to the dynamical time,  and the
flow can be characterized as gravitational infall.  Inward of $3\mpc$,
pressure forces begin to affect  the flow, and the flow velocity drops
below  the  dynamical  velocity.   At  $r\sim 0.5  \mpc$,  the  radial
velocity of  the flow is only  $\sim 50\kms$, and  inward of $0.5\mpc$
$\left|V_r\right|$ continues to  decrease.  In Figure~\ref{radial3} we
plot the  spherically averaged radial  velocity of the gas  divided by
the  local adiabatic  sound speed,  i.e.  the  Mach number.   The flow
undergoes  a  relatively  rapid  transition from  subsonic  to  mildly
supersonic flow in the region $2.5\mpc < r < 3.5\mpc$, suggesting that
a weak accretion shock is in  this region.  This places the shock at a
slightly  larger radius  than  that implied  by  the rise  in the  gas
density profiles.   The main differences  between the results  for the
\lowr\  simulation and  the other  two simulations  are caused  by the
smaller sound  speed in  the \lowr\ simulation.   In the  very central
regions, the \lowr\ simulation appears to have a radial outflow.  This
feature is well below the spatial resolution of this simulation and is
a numerical artifact.

At $\sim1\Mpc$  the radial  velocity of the  dark matter  component in
each  of the simulations  has a  strong dip,  indicating a  more rapid
infall of  material at  this radius caused  by a significant  clump of
material currently merging with the cluster. Closely examining the gas
radial infall plots, the gas in both the \lowr\ and \adia\ simulations
has similarly enhanced infall velocities at this radius, implying that
the infalling clump contains both  dark matter and gas.  This feature,
however, is not  present in the gas of the  \clus\ simulation, as most
of the gas  in the merging clump has collapsed  and turned into stars.
Such infalling clumps, coupled  with the action of dynamical friction,
could  represent a  primary  mode  of transport  of  baryons into  the
central regions of the cluster.

What causes  the differences in  the gas temperature  profiles between
the \clus\ and \adia\  simulations?  We conjecture that two processes,
either singly or jointly,  could be responsible for these differences.
First,  supernova feedback  could  have heated  the  gas.  The  radial
profile  of the cumulative  baryon fraction  (Figure~\ref{bar3}) shows
that a non-negligible fraction of the baryons is in the form of stars.
As a result of long local  cooling time scales, the energy injected by
the  associated supernova explosions  might not  be radiated  away and
might heat the  gas.  The second possibility is  an indirect impact of
cooling.  Cooling allows  some of the gas to  collapse and form stars,
removing   its  contribution   from  the   pressure  support   of  the
intracluster gas.  Much of this collapsed, stellar component forms the
dense, dynamically  significant object in  the center of  the cluster.
The gravity from this object draws in both the dark matter and the gas
in the  cluster; the drawing  in of the  former manifests itself  as a
steepening  in the  dark matter  density  profile.  The  gas would  be
similarly  drawn   in  but  unlike   the  dark  matter   would  suffer
compressional  heating.   Due to  the  long  cooling  time scale,  the
resultant thermal energy would not  be radiated away.  There is cooler
gas in the central object, as  well as in other objects that one would
identify as galaxies,  but the scale of these  objects is much smaller
than the scale of the cluster, and cold, dense gas turns into stars.

The radial profiles of the  entropy, $S(r)$, for the \adia\ and \clus\
simulations  (see  Figure~\ref{entropy3})   suggests  that  the  major
differences in  the gas density  and temperature profiles for  the two
clusters  (outside  the  very  central  core  region)  are  caused  by
compressional heating  of the gas  being drawn adiabatically  into the
deep central potential rather than  by supernova heating.  It is as if
a given  mass (gas)  shell at some  radius R$_1$ and  entropy $S\Delta
M_g$ in the  \adia\ simulation has simply been moved  to a radius R$_2
<$ R$_1$ in the \clus\ simulation.   To verify that this is indeed the
case,  we determined  the mean  entropy  per unit  mass [$\int  S(r)\;
dM_g(r) / M_g(r<r_{vir})$] between the clustercentric radius R and the
virial radius, as a function of the gas mass in the same volume.  This
quantity is unaffected by star formation.  Under the above hypothesis,
the  results   for  the  \clus\  and  \adia\   simulations  should  be
indistinguishable.   As  shown  in Figure~\ref{massentropy3}  this  is
indeed the  case, and  the results for  the \lowr\ simulation  are the
same.  Since the mean entropy per unit mass of the intracluster gas in
all  three simulations  is essentially  the same,  supernova feedback,
present only  in \clus\ simulation, cannot  be significantly affecting
the  intracluster medium.   The differences  in the  gas  density, gas
temperature,  and   specific  entropy  profiles  are   caused  by  the
differences  in the  nature  of the  potential  wells and  by the  gas
responding adiabatically to establish hydrostatic equilibrium in these
potentials.  Hence one  can understand the distribution of  gas in the
various  simulations in  terms of  individual shells  of  gas behaving
adiabatically.  This  does not  imply, however, that  the intracluster
gas has  a uniform entropy  distribution (e.g.~Figure~\ref{entropy3}).
The  entropy profile  depends on  the  accretion history  of the  gas,
including the strength of shocks.

The  gas distributions within  the simulations  presented here  can be
described  by  a  polytropic  equation of  state  with  $\gamma\approx
1.1$-$1.2$.  Such a conclusion was also reached for a sample of nearby
X-ray  emitting  clusters using  ASCA  observations \citep{ma98}.   In
their study of cluster  mass estimates using gravitational lensing and
X-ray  methods,  Miralda-Escud\'{e} \&  Babul  (1995)  found that  the
assumption that the intracluster gas  is isothermal (as was thought at
that time)  results in a  discrepancy between the two  mass estimates,
while the  assumption that the gas was  polytropic with $\gamma\approx
1.1$   resolves  the  discrepancy.    The  similarity   between  these
observational  results  and   our  simulations  is  encouraging.   The
polytropic nature of  the gas appears to be a  robust result, since it
occurs  in all  three simulations  even  though they  differ in  their
resolution and  in the included  physical processes.  We  will further
investigate the origin of such a distribution in our next paper, where
we study the physics of  cluster formation by examining the history of
the  dark  matter  and  gas  properties  through  the  course  of  the
simulation.

Unlike  simple  gravitational  physics  and the  hydrodynamics  of  an
adiabatic gas, the behavior of a  gas that cools and forms stars could
depend  on the prescription  one uses  to model  these processes  in a
simulation.  Although  the prescription used in  the \clus\ simulation
may not  be a completely  accurate description of star  formation, the
analysis presented  here shows that cooling and  star formation cannot
be neglected.  Consider, for example,  the X-ray flux from a simulated
cluster. The X-ray  luminosity for the \clus\ simulation  is ${\rm L_x
\sim 3\times10^{43} erg\ s^{-1}}$, about three times that measured for
the  Virgo cluster  of galaxies  (which the  simulation was  chosen to
represent).     This   contrasts   with    the   simulations    of   a
$7\times10^{14}\msun$  cluster by Suginohara  \& Ostriker  (1998), who
find  an X-ray  flux  of $7.4\times10^{45}{\rm  erg\  s^{-1}}$ with  a
temperature  of  only  3  keV.   Clusters with  this  temperature  are
observed to  have a  much smaller flux,  by more  than a factor  of 10
\citep{da93}.  While the Suginohara \& Ostriker (1998) simulations and
the \clus\  simulation presented here  both allow gas to  cool, theirs
did not  include star formation.   Star formation turns  cooler, dense
gas into stars  and removes it as a significant  source of X-rays, and
it appears that  this additional physics goes a  long way to repairing
the discrepancy between predicted  and observed X-ray properties found
by  Suginohara  \&  Ostriker  (1998).   Therefore we  do  not  require
additional  heating via  conduction or  an excessively  high supernova
rate  to reproduce  observed X-ray  luminosities.  However,  while the
reasonable X-ray properties of our simulated clusters are encouraging,
it  is  also clear  that  these properties  depend  in  detail on  the
resolution and  physics of the simulation.  In  the \clus\ simulation,
the reasonable X-ray luminosity  has been achieved through the removal
of  some of  the X-ray  emitting gas  by conversion  into  stars.  The
fraction of  baryons in stellar form,  however, is 30\%  as opposed to
the observed $\sim$10\% and even  then, \clus\ cluster has the largest
total X-ray luminosity.

In previous  simulations of clusters  that have included  cooling, but
not the  subsequent formation  of stars, the  pooling of gas  into the
cluster core has  resulted in the formation of  a bright central X-ray
peak  approximately  $200  \kpc$   in  extent  (Katz  \&  White  1993;
Suginohara \&  Ostriker 1998).  As demonstrated here,  the addition of
star  formation and  supernova  feedback into  the baryonic  component
reduces the  brightness and the  size of the  X-ray cusp but  does not
eliminate its  formation.  In  our simulation, the  X-ray cusp  is the
result of stellar  baryons forming a central object  massive enough to
influence the dark matter profile  significantly.  On the basis of the
X-ray  surface brightness  profile,  the \clus\  cluster most  closely
resembles  the XD  clusters studied  extensively by  Jones  and Forman
(1984).  X-ray  cusps are  seen in $\sim  70\%$ of  observed clusters,
with  the enhanced  central emission  interpreted  as being  due to  a
cooling flow \citep{pe98}.  While the \clus\ simulation does exhibit a
central  X-ray cusp  as well  as evidence  for a  cooling flow  in the
central  $40  \kpc$,  the  flow  has  no  significant  impact  on  the
kinematics beyond $40 \kpc$.

Even  with the  inclusion  of cooling  and  star formation,  simulated
clusters of  galaxies may not  truly resemble `real'  galaxy clusters.
However,  this does not  imply that  the complex  processes associated
with baryonic matter  can be neglected in numerical  studies of galaxy
clusters.   If simulations  with  cooling and  star  formation do  not
reproduce  the  observed properties  of  galaxy  clusters,  it may  be
because a  much more sophisticated  treatment of cooling  is required,
both in  physical modeling or numerical  treatment, or it  may be that
the simulations neglect other important processes like magnetic fields
permeating the  intracluster medium or  feedback from AGN.   A similar
conclusion was also reached by Suginohara \& Ostriker (1998).

\section{Conclusions}
\begin{itemize}
\item Moore et al. (1998) demonstrated that changing the resolution of
a simulation results  in a change in the slope  of the density profile
within  the  spatial  resolution  scale.   We  confirm  this  for  the
simulations presented here and similarly find that the central density
profile is approximately $\rho  \propto r^{-1.4}$.  With the inclusion
of cooling and  star formation, however, the form  of the mass profile
is  radically  different,  having  a  very steep  central  cusp.   The
combination of this deeper central potential well and the reduction in
gas density  in the  cluster's outer parts  (because of  conversion to
stars)  affects   many  of  the  cluster's   physical  and  observable
properties. Also,  we find that  our lower resolution  simulation does
not properly  capture the  central density profile,  illustrating that
sufficient resolution is required to obtain convergent results.

\item The  scaling of the  three-dimensional gas temperature  with gas
density is  similar to that deduced by  Markevitch~et~al.~(1998) for a
sample of nearby X-ray clusters. This result appears to be independent
of  the details  of  the  simulation.  The  gradient  in the  observed
temperature profile depends  on projection effects (as well  as on the
resolution of the X-ray telescope).

\item The X-ray surface brightness profile and total luminosity depend
sensitively on  both resolution and  physics. In the  simulation where
gas cools and forms stars, a bright central X-ray cusp develops in the
X-ray  surface  brightness profile.   The  total  X-ray luminosity  is
reasonable for a Virgo-like  cluster; however, the fraction of baryons
in the form of stars is 30\%, which is higher than typical fraction in
observed clusters  by a factor of  $\sim 3$.  Central  X-ray cusps are
observed in ${\rm \sim70\%}$ of galaxy clusters and are interpreted as
being due to cooling flows  \citep{pe98}.  While a flow is apparent in
the \clus\ simulation, its scale $({\rm \sim 40\kpc)}$ is smaller than
those  typically   seen  in  observed  galaxy   clusters  $({\rm  \sim
150\kpc)}$.  Based on the X-ray surface brightness profile, the \clus\
cluster most closely resembles an XD cluster.

\item
The  fitting of  $\beta$-models,  a common  practice  in cluster  mass
determinations from X-ray data, is  sensitive to the radial range over
which the fit is made.  In our simulations, the best fitting model out
to  beyond the  virial radius  has $\beta\sim1$,  but a  fit  over the
restricted  radial  range  typically observed  yields  $\beta\sim0.7$.
This suggests  that the observed discrepancy  between $\beta$ measured
from   cluster  X-ray  surface   brightness  distributions   and  that
determined  from  X-ray  temperatures  and  cluster  dynamics  may  be
resolved  if  the  fit   to  the  observed  X-ray  surface  brightness
profile could be extended to large radii.

\item The  gravitational lensing characteristics of  a galaxy cluster,
such  as  the  radial positions  and  widths  of  giant arcs  and  the
distribution  of weakly  lensed  images, depend  on  the core  radius,
central  value,   and  slope   of  the  projected   mass  distribution
(e.g.~Hammer 1991), rather  than on the total cluster  mass.  We find,
due to  the non-spherical nature  of the clusters in  our simulations,
that these properties also  depend on cluster orientation with respect
to  an  observer.   Hence,  the  mass distribution  inferred  from  an
analysis  of  the  gravitational  lensing characteristics  would  also
depend on  the viewing  angle.  The X-ray  profile, however,  does not
depend strongly  on the viewing  angle.  This might help  to partially
explain the  observed discrepancy between cluster  masses derived from
gravitational lensing techniques and those derived from X-ray analyses
(e.g~Allen 1998).  The  increased central concentration in simulations
with radiative cooling and star formation may rescue some cosmological
models  that otherwise  fail to  reproduce the  observed  incidence of
cluster arcs (cf. Bartelmann et al.\ 1998).

\item  Due  to   computational  limitations,  simulations  of  cluster
formation that have included a baryonic component have usually treated
it as  a simple adiabatic  gas. Given that  gas in clusters  must have
undergone  cooling,  collapse  and  processing  into  stars,  is  this
assumption  of  adiabaticity  valid?   Within  our  prescription  both
cooling and  star formation proceed rapidly, resulting  in very little
cold  gas pervading the  cluster environment.   The remaining  hot gas
continues to behave as an adiabatic fluid (Figure~\ref{massentropy3}).
However, ignoring these processes does  not result in the same cluster
evolution.   Significant differences exist  in all  cluster properties
for the \adia\  and \clus\ simulations.  In the  latter, as gas enters
the forming  cluster its  density rises, so  it cools  efficiently and
turns into  stars.  The reduction  in pressure support in  the central
regions causes more  gas to pool into the  center and undergo cooling.
This  modification of  pressure support,  coupled with  the increasing
central  density caused  by  the accumulation  of stars,  dramatically
influences many of the cluster  properties on scales out to the virial
radius.   It   is  important  to  note,  therefore,   that  while  the
prescription used in the simulations presented here may not completely
describe the cooling and subsequent star formation in galaxy clusters,
the effects  of these processes cannot  be neglected if  one wishes to
draw physical inferences from numerical simulations.
\end{itemize}

\section{Acknowledgments}
We would like to thank J. M. Gelb and E. Bertschinger for their N-body
simulations, and  J.  Raymond  for discussions on  the details  of his
plasma code.  We would also like  to thank George Lake and Eric Linder
for useful  discussions and the anonymous referee  for useful comments
and suggestions that improved the clarity of the paper.  This work was
supported in part by the NCSA and SDSC supercomputing centers, by NASA
Astrophysics Theory Grants NAGW-2422, NAGW-2523, NAG5-3111, NAG5-3820,
NAG5-4242,  and  NAG5-4064, by  NASA  LTSA  grant  NAG5-3525, by  NASA
HPCC/ESS grant NAG 5-2213, and by the NSF under grant ASC 93-18185, as
well an NSERC Research grant.  GFL acknowledges support from a Pacific
Institute  of Mathematical Sciences  Fellowship.  AB  acknowledges the
hospitality of the University of Washington.

\newpage
\begin{figure*}
\centerline{
}
\caption[]{The radial  dependence of the  density of all  matter, both
the dark and baryonic components, in each of the simulations presented
in  this  paper.  The  bars  along  the  x-axis indicate  the  spatial
resolution  for  the  medium  and  low  resolution  simulations.   The
equivalent bar for  \hire\ lies off the left of  the figure. The heavy
bars that cross the individual curves indicate the radius within which
each cluster has only 32 particles.}
\label{total3}
\end{figure*}

\begin{figure*}
\centerline{
}
\caption[]{The    difference     between    the    distributions    in
Figure~\ref{total3} and the model given by Equation~\ref{navarro}.}
\label{diff3}
\end{figure*}

\begin{figure*}
\centerline{
}
\caption[]{  As  in   Figure~\ref{total3},  but  plotting  the  radial
dependence of only the dark matter density in each of the simulations.
Again, the  bars along the  x-axis indicate the softening  lengths for
the  medium  and low  resolution  simulations,  while  the heavy  bars
indicate the radius within which the cluster has only 32 particles.}
\label{dark3}
\end{figure*}

\begin{figure*}
\centerline{
}
\caption[]{The radial dependence for  the circular velocity in each of
the  cluster simulations.   The  bars along  the  x-axis indicate  the
spatial resolution for the medium and low resolution simulations.  The
thick solid line in this figure  shows the circular velocity of a halo
with an NFW profile. }
\label{vcirc3}
\end{figure*}

\begin{figure*}
\centerline{
}
\caption[]{The radial  dependence of  the gas density  in each  of the
simulations.  The bars along the x-axis indicate the softening lengths
for  the medium  and low  resolution  simulations and  the heavy  bars
crossing the  curves indicate the  radius within which  the simulation
has only 32 particles.}
\label{gas3}
\end{figure*}

\begin{figure*}
\centerline{
}
\caption[]{The radial dependence of  the cumulative baryon fraction in
each  of the  simulations. The  two solid  lines represent  the \clus\
simulation: the total baryonic content of stars and gas (top line) and
only gas (bottom  line). The bars along the  upper x-axis indicate the
spatial resolutions of the  low and medium resolution simulations. The
central cusp in  the \lowr\ simulation is well  within the 32 particle
resolution limit.}
\label{bar3}
\end{figure*}

\begin{figure*}
\centerline{
}
\caption[]{The distribution of the  various baryonic components in the
\clus\  simulation. Here,  the density  is  in units  of the  critical
density and  the black line  represents the total baryonic  mass.  The
central peak is  comprised mainly of stars (dark  grey line), although
hot gas  (lighter grey line)  becomes the dominant  baryonic component
beyond $\sim130\kpc$.   Other than  a peak at  the very center  of the
cluster,  there  is  very   little  cold  (${\rm  T<10^{6.5}K}$)  gas,
represented by the lightest grey  line, in the cluster.  This has been
converted into stars.}
\label{colddense}
\end{figure*}

\begin{figure*}
\centerline{
}
\caption[]{The  cooling time  scale for  the intracluster  gas  in the
three simulations  with baryonic component.  Note the  change of scale
from logarithmic to linear at  1Mpc.  The sharp spikes apparent in the
\clus\ simulation are knots of  cold, collapsed gas. This gas is being
processed into stars and represents galaxies.}
\label{tcool3}
\end{figure*}

\begin{figure*}
\centerline{
}
\caption[]{  As   in  Figure~\ref{total3}  but   plotting  the  radial
dependence of the gas temperature  in each of the simulations. The bar
in the upper  left-hand corner indicates the softening  length for the
low  resolution  simulation.  The  horizontal  line  shows the  virial
temperature of the cluster ${\rm (T_{vir}\sim2.8\ keV)}$.  }
\label{temp3}
\end{figure*}

\begin{figure*}
\centerline{
}
\caption[]{   As  in   Figure~\ref{total3},   presenting  the   radial
dependence of the gas entropy  (top) and dark matter (bottom) per unit
mass in each  of the simulations.  The bars  along the x-axis indicate
the  softening length  for the  low and  high  resolution simulations.
Note the change in scale from logarithmic to linear at 1Mpc.}
\label{entropy3}
\end{figure*}

\begin{figure*}
\centerline{
}
\caption[]{  A   comparison  of  the  gas  and   dark  matter  entropy
distributions for each of the simulations with gas.  Each presents the
gas (thin line) and dark matter (thick line) entropy for \clus, \adia\
and \lowr\ simulations.  }
\label{darkentropy3}
\end{figure*}

\begin{figure*}
\centerline{
}
\caption[]{ Each panel plots the projected surface mass density versus
radius.  The left-hand panels plot  the total mass in the simulations,
while the right-hand  panels plot only the dark  matter.  The {\bf X},
{\bf Y}  and {\bf Z}'s denote  the projection axes,  which are aligned
with the coordinate  axes of the simulation.  The  overdensity is with
respect to integrating a uniform density of ${\rm \Omega = 1}$ through
a  line-of-sight  of  4Mpc, the  size  of  the  box from  which  these
projections were made.  }
\label{mass2}
\end{figure*}

\begin{figure*}
\centerline{
}
\caption[]{The projected X-ray emission: the left-hand panel shows the
flux radiated into 4$\pi$  steradians in the $2\rightarrow10$ keV band
and   the  right-hand   panel  shows   the  flux   in   the  ROSAT/HRI
$(0.1\rightarrow2.4{\rm\ keV})$ band.   This band is ``windowed'' with
the normalized response function of the HRI instrument.  }
\label{xray2}
\end{figure*}

\begin{figure*}
\centerline{
}
\caption[]{The 2$\rightarrow$10 keV  X-ray surface brightness profiles
for  the \clus\  (left)  and \adia\  (right)  simulations; each  panel
displays the  \X, \Y\  and \Z\ cluster  projections.  The  thick solid
line in  each panel plots the best  fit $\beta$-model (eq.~\ref{beta})
over  a restricted  range from  50  kpc to  750 kpc,  similar to  that
employed with  real X-ray data,  while the dot-dashed lines  plots the
best fit $\beta$-model over the full  range of the data.  To avoid the
effects of the X-ray peaks in  the cluster center and at the positions
of galaxies on the mean profile  of the \clus\ simulation, we use only
the  circled  points in  the  fits.  The  best  fit  values for  ${\rm
(\beta,r_c)}$ for the \clus\  simulation are (0.97, 218kpc) and (0.63,
72kpc)   for  the   full  and   restricted  range   respectively.  The
corresponding values for the  \adia\ simulation are (1.07, 211kpc) and
(0.77, 106kpc).}
\label{beta2}
\end{figure*}

\begin{figure*}
\centerline{
}
\caption[]{ Like Figure~\ref{xray2} but plotting the projected X-ray
emission weighted temperature profiles.  }
\label{temp2}
\end{figure*}

\begin{figure*}
\centerline{ 
}
\caption[]{The  projected  \SZ\  decrement  for  the  various  cluster
simulations.}
\label{sz2}
\end{figure*}

\begin{figure*}
\centerline{ 
}
\caption[]{The  mass accretion  rate of  gas through  spherical shells
into  the  cluster  normalized  by  the  accretion  rate  required  to
accumulate  the mass  of gas  within the  cluster virial  mass  at the
present time ${\rm ( 2.05\times10^{13}\msun)}$ in a Hubble time.  Note
the change of scale from logarithmic to linear at 1Mpc.}
\label{mass_accretion3}
\end{figure*}

\begin{figure*}
\centerline{
}
\caption[]{The spherically averaged radial  infall velocity of the gas
divided by the local dynamical velocity,  which we define as ${\rm r /
t_{dyn}}$.  Note  the change  of scale from  logarithmic to  linear at
1Mpc. }
\label{dynamic3}
\end{figure*}

\clearpage

\begin{figure*}
\centerline{
}
\caption[]{The spherically averaged radial infall velocity of the gas
in  units of the   local sound  speed.  Beyond the  virial
radius a weak shock is present.  Note the change of scale
from logarithmic to linear at 1Mpc.}
\label{radial3}
\end{figure*}

\begin{figure*}
\centerline{
}
\caption[]{The total entropy in the cluster as  a function of gas mass
between a radius r and the virial radius. Both axes have been normalized
by the mass of gas within the virial radius.}
\label{massentropy3}
\end{figure*}

\clearpage

\newcommand{\pppa}{a}
\newcommand{\pppb}{b}
\newcommand{\pppc}{c}
\newcommand{\pppd}{d}
\newcommand{\pppe}{e}
\newcommand{\pppf}{f}
\newcommand{\pppg}{g}
\newcommand{\ppph}{h}
\newcommand{\pppi}{i}
\newcommand{\pppj}{j}
\newcommand{\pppk}{k}
\newcommand{\pppl}{l}
\newcommand{\pppm}{m}
\newcommand{\pppn}{n}
\newcommand{\pppo}{o}

\newcommand{\xx}{}

\begin{deluxetable}{ccccccc}
\tablecolumns{7}  
\tabletypesize{\scriptsize}
\tablecaption{A summary of previous numerical studies of galaxy clusters. 
\label{table1}}
\tablehead{
\colhead{\xx Ref.} & \colhead{\xx Spatial ($h^{-1}$kpc)} &  
\colhead{\xx Dark($h^{-1}{\msun}$) } &
\colhead{\xx Virial ($h^{-1}{\msun}$)}    & \colhead{\xx Gas($h^{-1}{\msun}$) } &
\colhead{\xx Cool / Star} & \colhead{\xx ${\rm M_{clust}}$($h^{-1}{\msun}$) }
}
\startdata
(\pppa)&  150 &3.0$\times10^{13}$&5.6$\times10^{12}$&3.4$\times10^{12}$& N &   $1.2\times 10^{15} (<R_A)$     \\
(\pppb)&  14 &4.8$\times10^{11}$&4.5$\times10^{9 }$&4.8$\times10^{10}$&Y/N &   1.0$\times10^{14}(<R_{200})$\\
(\pppc)&270$h$ p&1.1$\times10^{14}h$&3.3$\times10^{13}h^3$&2.2$\times10^{11}h$&N&           1.85$\times10^{15}h (<R_{200})$\\
(\pppd)& 40 &3.8$\times10^{12}$&1.1$\times10^{11}$&4.2$\times10^{11}$& Y/N &              5.9$\times10^{14}(<R_{200})$\\
(\pppe)&628 &1.2$\times10^{11}$&4.1$\times10^{14}$&1.2$\times10^{10}$&    N &           many                 \\
       &    &                  &                  &(2.1$\times10^{12}$)&       &                                \\
(\pppf)&  78
&2.0$\times10^{12}$&7.9$\times10^{11}$&2.2$\times10^{11}$& N    &
$4.7\times 10^{14} (<R_{400})$\\
(\pppg)&  50 p&9.4$\times10^{11}$&2.1$\times10^{11}$&9.4$\times10^{10}$& N    &            $2.5\times 10^{14}(<R_{200})$\\
(\ppph)& 50 &4.8$\times10^{12}$&2.1$\times10^{11}$&2.6$\times10^{11}$&N &          3-20 $\times 10^{14}(<R_{500})$\\
(\pppi)&  98 &1.6$\times10^{10}$&1.6$\times10^{12}$&1.7$\times10^{ 9}$& N &             many               \\
       &    &                  &                  &(3.0$\times10^{11}$)&       &                                \\
(\pppj)& 14 p  &2.3$\times10^{12}$&4.5$\times10^{ 9}$&2.6$\times10^{11}$&N &             1.0$\times10^{15}(<R_{200})$\\
(\pppk)&  66 &8.0$\times10^{11}$&4.8$\times10^{11}$&4.9$\times10^{11}$&Y/N &          $>1.2 \times 10^{14}(<R_A)$  \\
\enddata
\tablecomments{
All values are for clusters at z=0.
Physical, rather than comoving, resolution scales are denoted by a p; corresponding virial
mass scales as ${\rm (1+z)^3}$. The studies are:
(\pppa)~\cite{ev90};
(\pppb)~\cite{ka93};
(\pppc)~\cite{sc93,sm93};
(\pppd)~\cite{th92};
(\pppe)~\cite{co94};
(\pppf)~\cite{me94};
(\pppg)~\cite{na97};
(\ppph)~\cite{ba96};
(\pppi)~\cite{br98};
(\pppj)~\cite{ek98};
(\pppk)~\cite{su98}.
}
\end{deluxetable}

\clearpage

\begin{deluxetable}{cccccccc}
\tablecolumns{8}  
\tabletypesize{\scriptsize}
\tablecaption{A   summary of the various  resolutions of
the simulations presented in this paper. 
\label{table2}}
\rotate
\tablewidth{540pt}  \tablehead{   \colhead{\xx  Name}  &  \colhead{\xx
Spatial ($h^{-1}$kpc)} & \colhead{\xx Dark Matter ($h^{-1}{\msun}$)} &
\colhead{\xx    Virial   ($h^{-1}{\msun}$)}    &    \colhead{\xx   Gas
($h^{-1}{\msun}$)}   &   \colhead{Cool/Star}   &  \colhead{\xx   ${\rm
n_{dark}/n_{gas}/n_{star}}$}   &  \colhead{\xx   ${n_{timesteps}}$}  }
\startdata   \dark   &14  &3.7${\rm\times10^{11}}$&4.6${\rm\times10^{9
}}$&NO     &NO     &18134/0/0     &     2400     \\     \clus     &14
&3.5${\rm\times10^{11}}$&4.6${\rm\times10^{9               }}$&1.9${\rm
\times10^{10}}$&YES&18221/11582/6987&     2400    \\     \adia    &14
&3.5${\rm\times10^{11}}$&4.6${\rm\times10^{9               }}$&1.9${\rm
\times10^{10}}$&NO     &18290/17017/0     &     2400     \\     \lowr
&200&3.5${\rm\times10^{11}}$&1.3${\rm\times10^{13}}$&1.9${\rm
\times10^{10}}$&NO     &17884/17708/0     &     2400     \\     \hire
&1.4&1.4${\rm\times10^{10}}$&4.6${\rm\times10^{6       }}$&NO      &NO
&494486/0/0  &10000 \\  \hline  \\ \multicolumn{7}{c}{Cluster  Vital
Statistics:        ${\rm        M_{clus}=2.1\times10^{14}h^{-1}\msun,\
R_{vir}=1h^{-1}Mpc}$ \&  ${\rm V_{circ}(R_{vir})=1000\kms}$ }  \\ \\
\enddata    \tablecomments{    Quantities    are   defined    as    in
Table~\ref{table1}.  Although we quote $h^{-1}$ units to be consistent
with Table~\ref{table1}, remember that  $h=0.5$ and that the ``spatial
resolution'' is  {\it twice} the gravitational  softening length.  The
final column refers to the number of particles, and their type, within
the virial radius.}
\end{deluxetable}

\clearpage

\begin{deluxetable}{ccc}
\footnotesize \tablecaption{Cluster X-ray properties
\label{table3}}
\tablewidth{250pt}  \tablehead{  \colhead{Name}  & \colhead{${\rm  L_x
(10^{43}erg\   s^{-1})}$}  &   \colhead{$\left<T_x\right>$   (keV)}  }
\startdata \clus  & 3.14 & 4.41  \\ \adia &  2.58 & 3.43 \\  \lowr &
0.88  & 2.16  \\  \enddata  \tablecomments{ A  summary  of the  X-ray
properties,  in  the  2$\rightarrow$10\   keV  energy  band,  for  the
simulated clusters  presented in this  paper.  The temperature  is the
mean emission weighted temperature averaged over a circle of radius of
the virial radius (change this to twice or half the virial radius does
not  significantly alter either  the X-ray  flux or  emission weighted
temperature for the simulated cluster). The cluster virial temperature
is 2.8 keV.  }
\end{deluxetable}

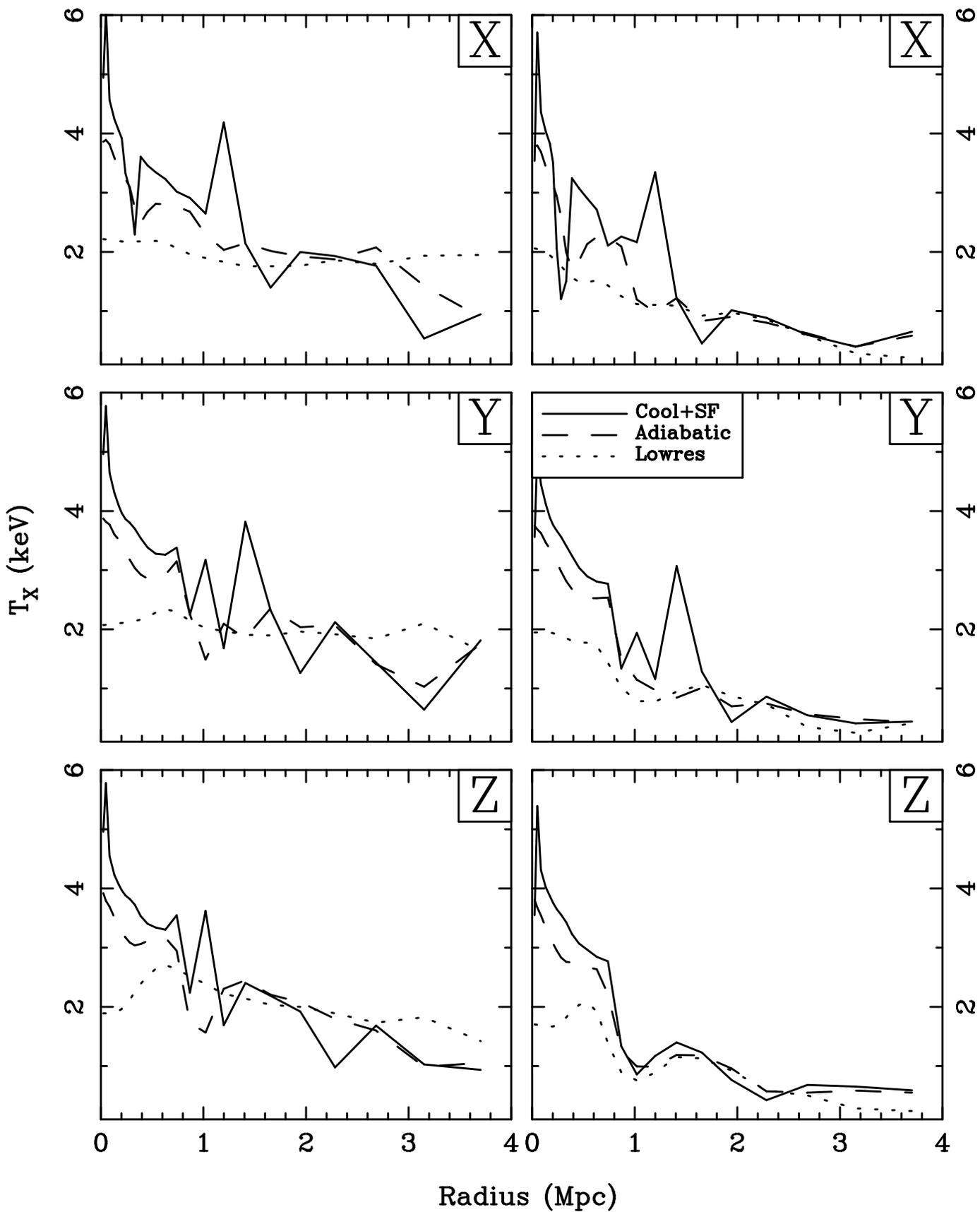
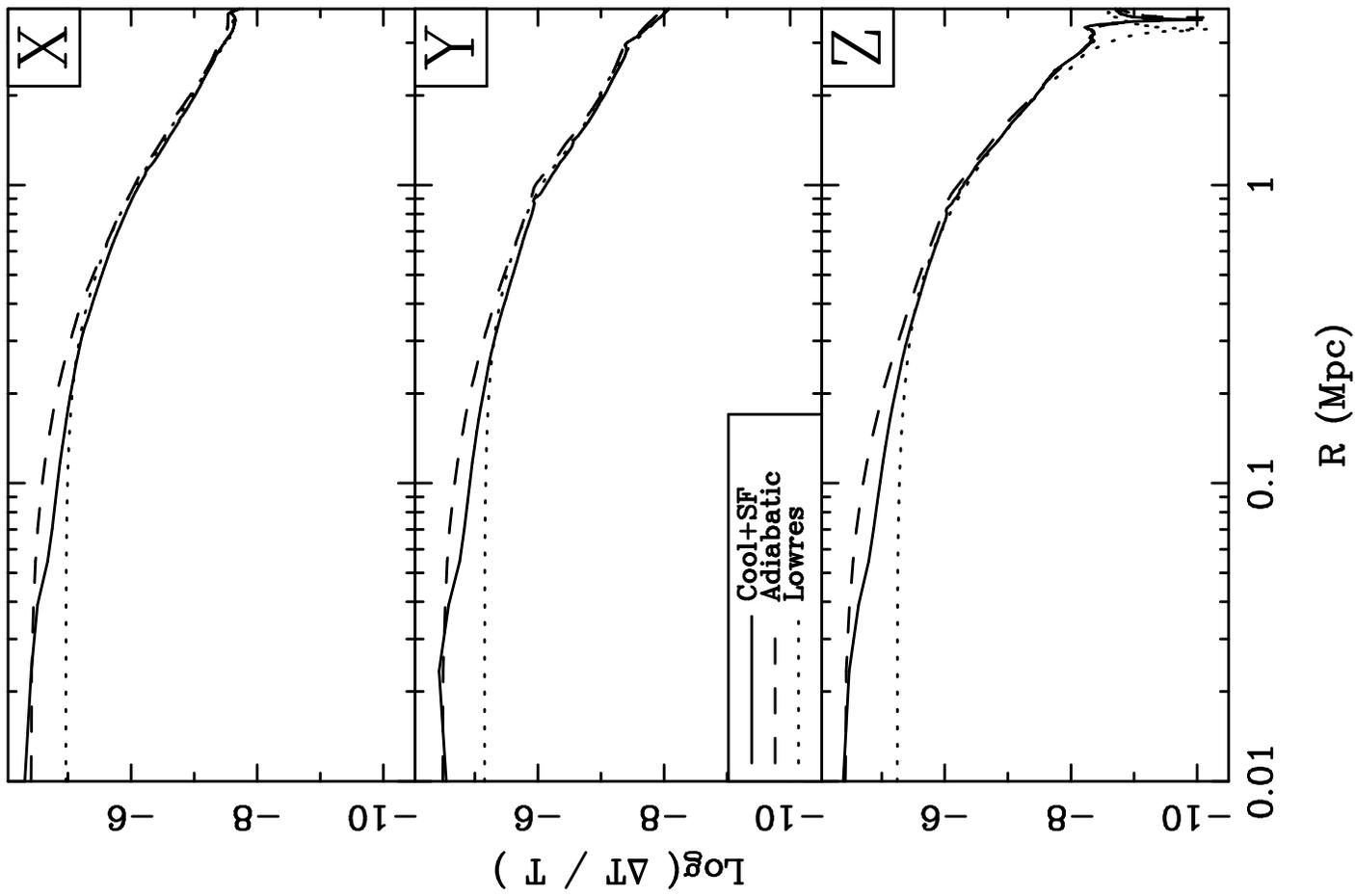
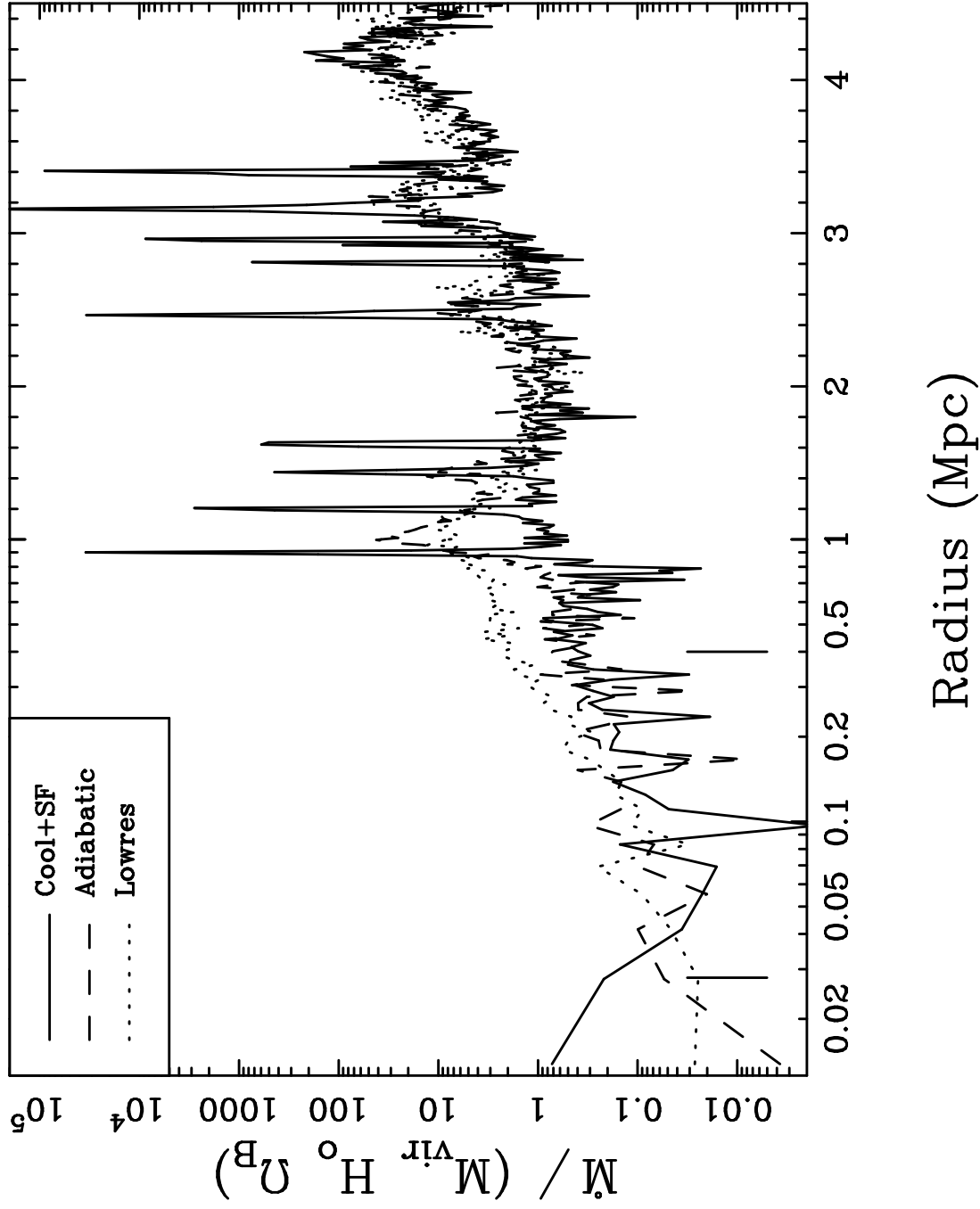
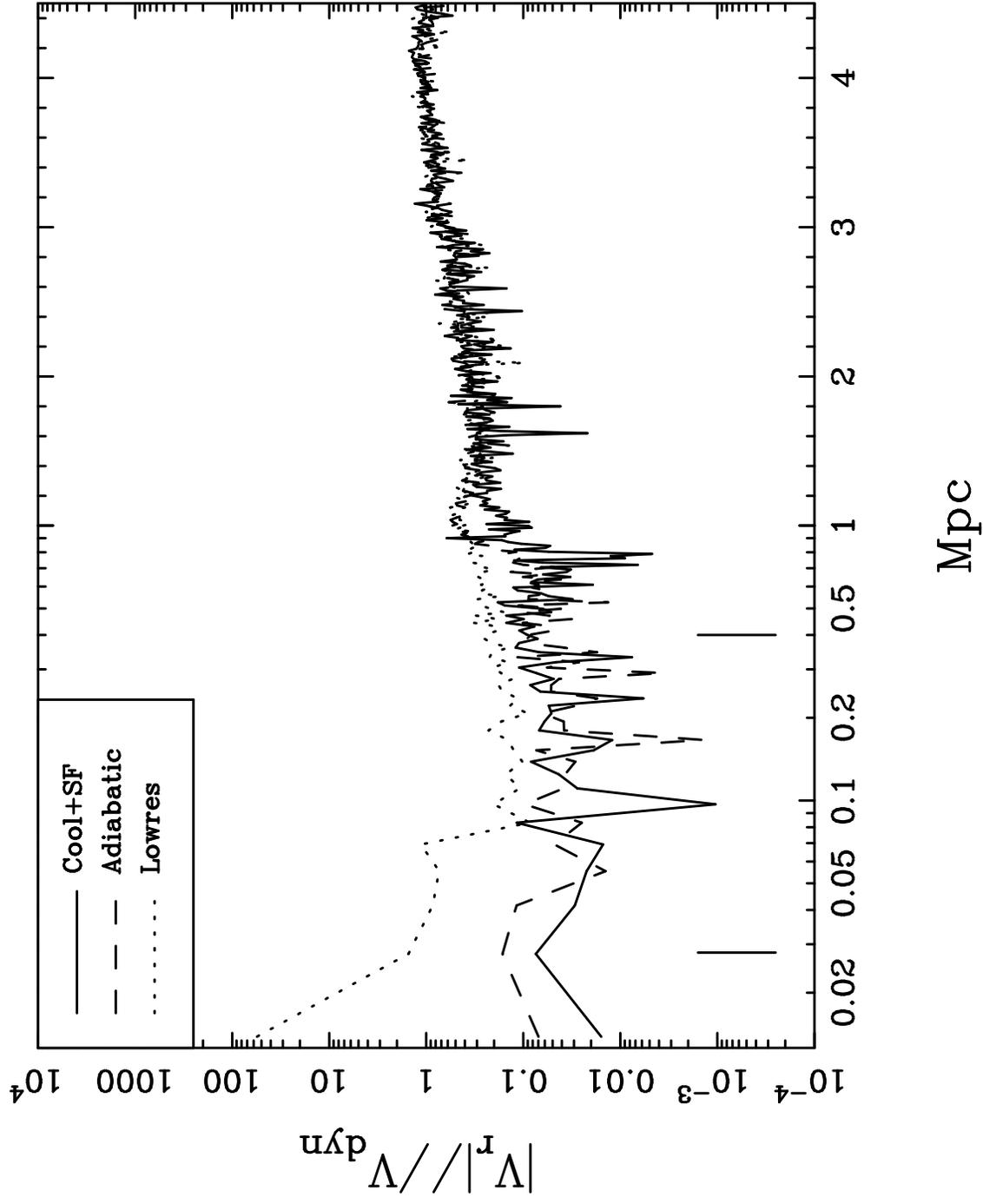
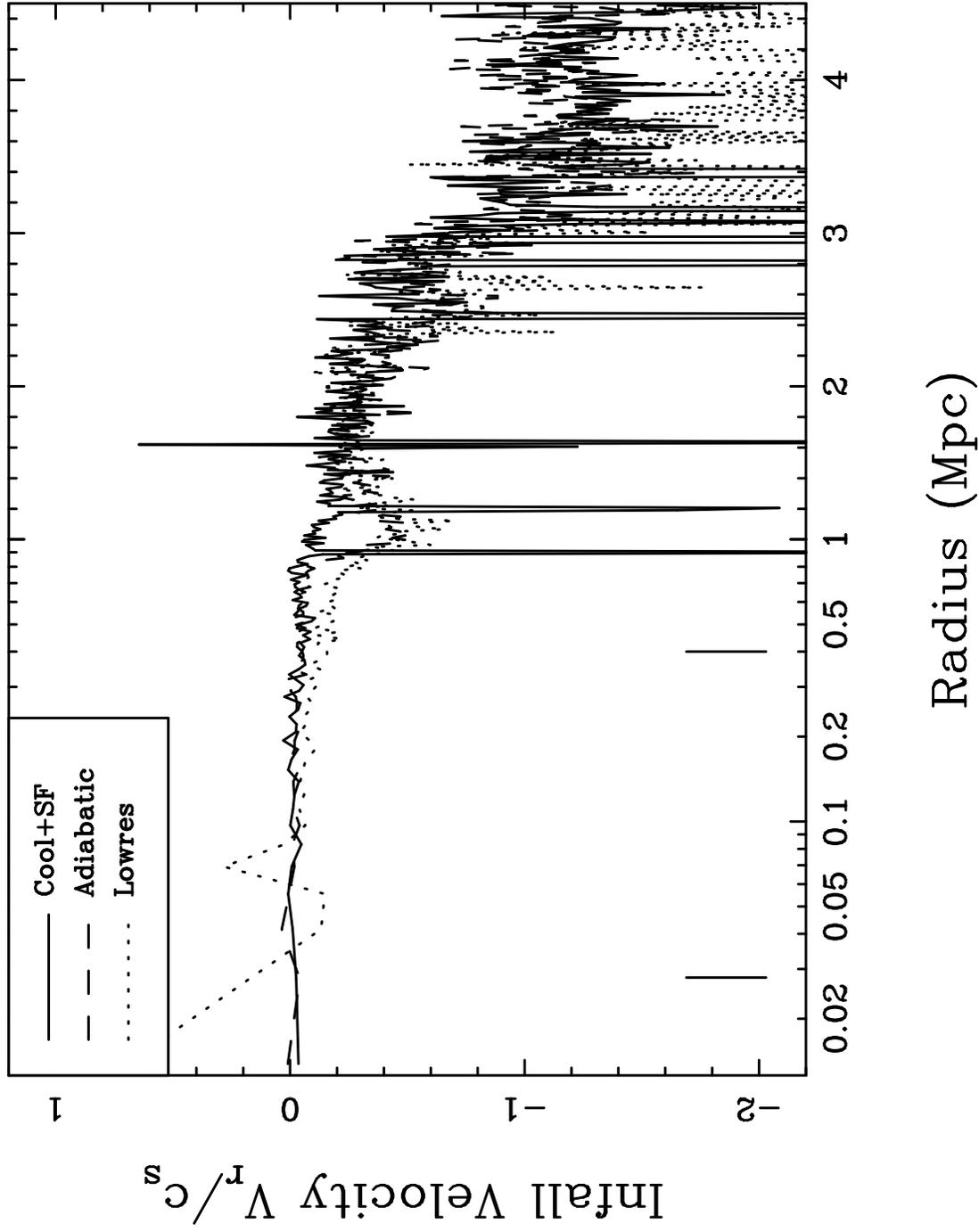
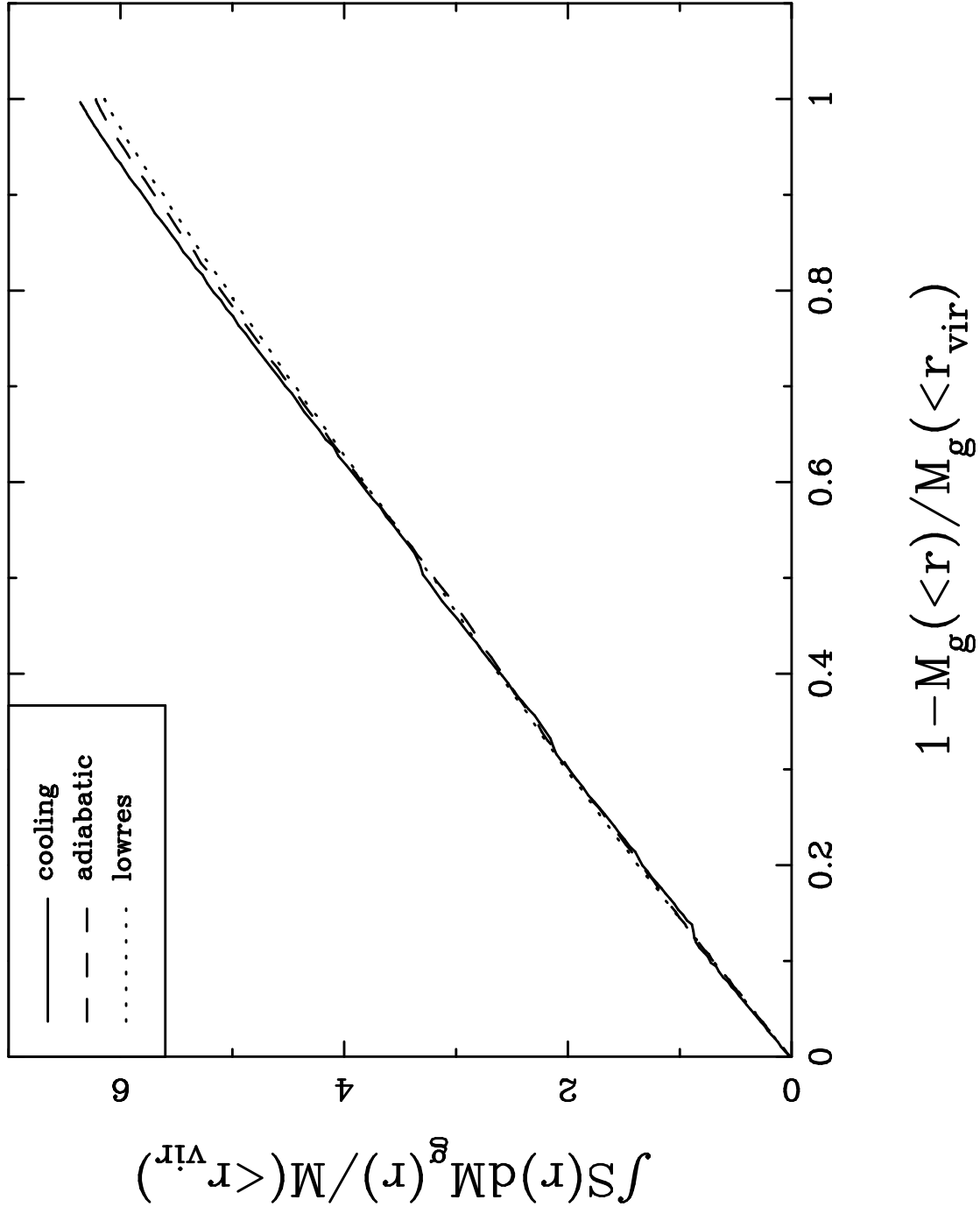
\end{document}